\pdfoutput=1 
\documentclass[letterpaper, 10 pt, conference]{ieeeconf}  % Comment this line out if you need a4paper

\IEEEoverridecommandlockouts                              % This command is only needed if 

\usepackage{amsmath} % assumes amsmath package installed
\usepackage{amssymb}  % assumes amsmath package installed

\usepackage{amsfonts}
\usepackage{amsmath}
\usepackage{mathrsfs}
\usepackage{mathtools}
\usepackage{soul,color}
\usepackage[dvipsnames]{xcolor}
\usepackage{comment}
\usepackage[normalem]{ulem}
\usepackage{amssymb,graphicx}
\newcommand\scalemath[2]{\scalebox{#1}{\mbox{\ensuremath{\displaystyle #2}}}}
\NewDocumentCommand{\tra}{om}{%
	\IfNoValueTF{#1}
	{#2}
	{#2_{[#1]}}%
}

\NewDocumentCommand{\pce}{om}{%
	\IfNoValueTF{#1}
	{\mathsf{#2}}
	{\mathsf{#2}^{#1}}%
}

\usepackage{algorithm}
\usepackage{algorithmicx}  
\usepackage{algpseudocode}
\usepackage{booktabs}
\renewcommand{\algorithmicrequire}{\textbf{Input: }}

\usepackage{adjustbox}
\usepackage{multirow}
\usepackage{siunitx}

\newcommand{\End}{\hfill $\square$}

\DeclareMathOperator*{\trace}{tr}
 \DeclareMathOperator*{\diag}{diag}

\newcommand{\mbb}[1]{\mathbb{#1 }}
\newcommand{\mbf}[1]{\mathbf{#1}} 
\newcommand{\mcl}[1]{\mathcal{#1}}

\newcommand{\norm}[1]{\|{#1}\|}

\newcommand{\pcecoe}[2]{\mathsf{#1}^{#2}}
\newcommand{\relx}{(\omega)}

\newcommand{\seq}[1]{\boldsymbol{#1}} %sequence

\newcommand{\trar}[2]{#1_{[0,{#2}]}}
\newcommand{\set}[1]{\mathbb{#1}}    

\newcommand{\inst}[1]{_{#1}}   
\newcommand{\pred}[2]{_{{#1}|{#2}}}

\newcommand{\splx}[1]{\mcl{L}^2(\Omega, \mathcal{F}, \mu; \mathbb{R}^{#1})}
\newcommand{\spl}{\mcl{L}^2(\Omega, \mathcal{F}, \mu; \mathbb{R})} 
 
\newcommand{\diff}{\mathop{}\!\mathrm{d}}

\newcommand{\mean}{\mbb{E}}

\newcommand{\prob}{\mbb{P}}
\newcommand{\hankel}{\mcl{H}}
 \newcommand{\covar}{\Sigma}

\newcommand{\Zf}{\mbb{Z}_{\text{f}}}

\newcommand{\rinv}{\dagger}
\newcommand{\Tini}{{T_{\text{ini}}}}
\newcommand{\lag}{\Tini}

\newcommand{\ini}{{\text{ini}}} % initial condition

\newcommand{\dimx}{{n_x}}
\newcommand{\dimy}{{n_y}}
\newcommand{\dimu}{{n_u}}
\newcommand{\dimw}{{n_w}}
\newcommand{\dimz}{{n_z}}
\newcommand{\dimv}{{n_v}}

\newcommand{\I}{\mathbb{I}}
\newcommand{\N}{\mathbb{N}}
\newcommand{\R}{\mathbb{R}}

\newtheorem{theorem}{Theorem}

\newtheorem{lemma}{Lemma}
\newtheorem{proposition}{Proposition}
\newtheorem{definition}{Definition}
\newtheorem{remark}{Remark}

\newtheorem{assumption}{Assumption}

\newcommand{\zdd}{ \boldsymbol{z}_\text{d}}
\newcommand{\ydd}{  \boldsymbol{y}_\text{d}}
\newcommand{\udd}{ \boldsymbol{u}_\text{d}}
\newcommand{\wdd}{ \boldsymbol{w}_\text{d}}

\newcommand{\Zkini}{ \bar{Z}_k}

\title{On  Data-Driven  Stochastic Output-Feedback Predictive Control}

\author{Guanru Pan, Ruchuan Ou and Timm Faulwasser$^{\star}$% <-this % stops a space
\thanks{$^{\star}$: Corresponding author.}%
\thanks{This work was supported by the German Research Foundation (Deutsche Forschungsgemeinschaft DFG) under Project 499435839.
	Major parts of this work were done while GP, RO, and TF were with the Institute for Energy Systems, Energy Efficiency and Energy Economics, TU Dortmund University, Dortmund, Germany. Currently, they are with the Institute of Control Systems at Hamburg University of Technology, Hamburg, Germany.
	{\tt\small $\{$guanru.pan,ruchuan.ou$\}$@tuhh.de timm.faulwasser@ieee.org} 
}
}

\begin{document}

\maketitle
\thispagestyle{empty}
\pagestyle{empty}

\begin{abstract}
 The fundamental lemma by Jan C. Willems and co-authors enables the representation of all input-output trajectories of a linear time-invariant system by measured input-output data. This result has proven to be pivotal for data-driven control.
Building on a stochastic variant of the fundamental lemma, this paper presents a data-driven output-feedback predictive control scheme for stochastic Linear Time-Invariant (LTI) systems. The considered LTI systems are subject to non-Gaussian disturbances about which only information about their first two moments is known. 
Leveraging polynomial chaos expansions, the proposed scheme is centered around a data-driven stochastic Optimal Control Problem (OCP). 	Through tailored online design of initial conditions, we provide sufficient conditions for the recursive feasibility of the proposed output-feedback scheme based on a data-driven design of the terminal ingredients of the OCP. Furthermore, we provide a robustness analysis of the closed-loop performance.  A numerical example illustrates the efficacy of the proposed scheme. 
\end{abstract}


\section{Introduction}
A frequently considered route to data-driven  control relies on the seminal insights of the fundamental lemma by Willems et al.~\cite{Willems2005}, see~\cite{DePersis19,Markovsky21r} for recent reviews. The lemma states that the set of all possible trajectories of a controllable LTI system can be characterized by measured input-output data, given that the input signal is persistently exciting. It serves as the basis for different data-driven controller design techniques. Specifically, it enables direct output-feedback predictive control that bypasses the intermediate identification step, cf. \cite{Yang15,Coulson2019,Berberich20}. We refer to \cite{Doerfler2023} for the comparison of direct data-driven methods and their indirect counterparts that utilize data for system identification. Applications in multiple domains are discussed in~\cite{Carlet2022,Bilgic22,Wang2022a}.
Sufficient conditions for recursive feasibility, stability, and robustness of output-feedback predictive control for deterministic LTI systems subject to noise-corrupted measurements are given in \cite{Berberich20,Berberich2021t,Bongard2022}.

Besides corrupting the measurement data, stochastic uncertainty can  directly enter the dynamics as process disturbance. To take this into account, \cite{Kerz21d} proposes a stochastic data-driven predictive control scheme based on probabilistic constraint tightening with closed-loop guarantees while requiring measurements of states. In \cite{Wang2022}, a data-driven output-feedback scheme based on the innovation form of the underlying system is considered, while the closed-loop properties remain unclear.
It is worth noting that in the aforementioned works, future inputs and outputs are only predicted in expectation, while in-depth statistical information about future disturbances---e.g., higher-order moments---are omitted or used only for constraint tightening. Moreover, in the case of model-based output feedback with Gaussian disturbance, the prediction of future statistical distributions can be achieved via the covariance dynamics of the state which in turn requires the design of a state estimator~\cite{Farina2015}.  

Our previous works~\cite{Pan21s,Faulwasser2022,Pan2023a} have made  first preparatory steps towards output-feedback data-driven stochastic predictive control. In \cite{Pan21s} we leverage Polynomial Chaos Expansions (PCE) to introduce a stochastic variant of Willems' fundamental lemma, enabling the data-driven forward propagation of statistical distributions for inputs and outputs.  In \cite{Faulwasser2022}, we give the related behavioral characterizations of stochastic LTI systems. Moreover,  we propose a \textit{state-feedback} predictive control scheme with closed-loop guarantees in \cite{Pan2023a}. Note that the latter work relies on a PCE basis which grows in dimension as the closed loop evolves.

The present paper addresses the open problem of closed-loop analysis for \textit{data-driven stochastic output-feedback} predictive control based on the stochastic fundamental lemma from \cite{Pan21s,Faulwasser2022}.
Our main contribution is an analysis that requires only limited information about the statistical properties of the stochastic disturbance, i.e., only its first two moments are assumed to be known. Specifically, we make the following contributions:
i)	we provide sufficient conditions for recursive feasibility of the proposed output-feedback scheme. Specifically, we consider the initial conditions of the  OCP solved in closed loop to be stochastic such that the predicted values from the last step can serve as feasibility-ensuring backups. Moreover, in contrast to \cite{Pan2023a}, which relies on a PCE basis whose dimension grows as the closed loop progresses,  here we work with a basis of constant dimension.
ii) we establish average asymptotic performance bounds through a data-driven design of the terminal ingredients.
iii) moreover,  we analyze the robustness of the performance of the closed loop. Instead of assuming exact measurements of past disturbances, our analysis considers the error resulting from the estimation of past disturbance realizations. 
iv)	crucially, the proposed scheme only requires information about the first two statistical moments of the disturbance instead of the exact knowledge of the underlying distribution. To this end, we tailor the distributionally robust uncertainty propagation scheme from~\cite{Pan2023}.

\subsubsection*{Notation}
    We denote as $(\Omega,\mathcal F,\mu)$ a probability space with sample space~$\Omega$,  $\sigma$-algebra~$\mathcal F$, and  probability measure~$\mu$.
	Furthermore,  $\splx{n_v}$  is the space of vector-valued random variables of dimension $n_v$ with realizations in $\R^{n_v}$,  finite expectation and covariance.
	Consider  a sequence of vector-valued random variables $V: \I_{[0,T-1]}$ $\rightarrow \splx{n_v}$. For an outcome $\omega \in \Omega$, we denote the mean, covariance, and realization of $V$ as $\mean[V]$, $\Sigma[V]$, and $v\doteq V(\omega)$, respectively. 
   The vectorization of $v : \I_{[0,T-1]}$ $\rightarrow \R^{n_v}$ is written as $\trar{v}{T-1} \doteq [v_0^\top,v_1^\top, \dots,v_{T-1}^\top]^\top \in \R^{n_v T}$. Similarly, we denote the vectorization of the sequence of random variables $V$  as $\trar{V}{T-1}$. Throughout the paper, we denote the identity matrix of size $l$ by $I_l$. 
    For a vector $x\in \R^{n}$, we consider $\|x\| \doteq \sqrt{x^\top x}$ and $\|x\|_Q \doteq \sqrt{ x^\top Q x}$, where $Q$ is positive semi-definite. As a shorthand for positive definite (positive semi-definite) matrices, we use $Q\succ 0$ ($Q \succeq 0$). 
 \begin{definition}\label{def:Hankel}
 	A Hankel matrix of depth $L \in \N^+$ associated with the signal sequence $v_{[0,T-1]}$ with $T \in \N^+$ is  \[
 	\hankel_L(\trar{v}{T-1}) \doteq \begin{bmatrix}
 		v\inst 0  &\cdots& v\inst{T-L} \\
 		\vdots & \ddots & \vdots \\
 		v\inst{L-1}& \cdots  & v\inst{T-1} \\
 	\end{bmatrix}.
 	\] 
 \end{definition}

\section{Preliminaries}\label{sec:preliminary}
To simplify later developments, we describe the input-output dynamics of LTI systems via the AutoRegressive with eXtra input (ARX) structure
\begin{equation}\label{eq:ARX}
	Y_{k} = \Phi Z_{k} + D U_{k} +W_{k},\quad Z_0 = \bar{Z}_{0}
\end{equation}
with input $U_k \in \mcl{L}^2(\Omega, \mathcal{F}_{k-1} , \mu; \mathbb{R}^{\dimu})$, output  $Y_k \in \mcl{L}^2(\Omega, \mathcal{F}_{k} , \mu; \mathbb{R}^{\dimy})$,  and disturbance $W_k \in \splx{\dimw}$ $(\dimw = \dimy)$. We use the shorthand notation
\begin{equation}\label{eq:extendedstate}
Z\inst{k} \doteq \begin{bmatrix}
	\tra{U}_{[k-\lag,k-1]}\\
	\tra{Y}_{[k-\lag,k-1]} \end{bmatrix}\in \mcl{L}^2(\Omega, \mathcal{F}_{k-1} , \mu; \mathbb{R}^{\dimz})
\end{equation}
 with $\dimz =\lag(\dimu+\dimy)$, i.e., $Z_k$ collects the last $\lag \in \N^+$ inputs and outputs. 
 We refer to $Z\inst{k}$ as the \textit{extended state variable}. In the underlying filtered probability space $(\Omega,\mcl F,(\mcl F_k)_{k\in \N}, \mu)$, the $\sigma$-algebra $\mcl F$ contains all available historical information, i.e, $\mcl F_0 \subseteq \mcl F_1 \subseteq \cdots \subseteq \mcl F$. Precisely, the filtration $(\mcl F_k)_{k\in \N}$ is defined by $\mcl F_0 = \sigma(Z_0)$ and $\mcl F_k = \sigma(Z_0, W_i, i \leq k)$, where $\sigma(Z_0, W_i, i \leq k)$ denotes the $\sigma$-algebra generated by $Z_0$ and $W_i, i \leq k$. The stochastic processes $U$, $Y$, and $Z$ are adapted to the filtration $(\Omega,\mcl F,(\mcl F_k)_{k\in \N})$, that is, $U_k$, $Y_{k-1}$, and $Z_k$ only depend on $Z_0$ and $W_i, i < k$. We refer to \cite{fristedt13modern} for details about filtrations.
 
 For a specific uncertainty outcome $\omega \in \Omega$, we denote the realization of $W_{k}$ as $w_{k} \doteq W_{k}(\omega)$. Likewise, the realizations of input, output, and extended state  are written as  
 \[u_k \doteq U_k\relx, \quad y_k \doteq Y_k\relx, \quad \text{and} \quad z_k \doteq Z_k\relx.\]
 Moreover,  given $\bar{z}_0 = \bar{Z}_0\relx \in \R^{\dimz}$ and $w_k$, $k\in \N $, the stochastic system \eqref{eq:ARX}  induces the \textit{realization dynamics}, i.e., the dynamics satisfied by path-wise sampled trajectories,  
 \begin{equation}\label{eq:ARX_realization}
 	y_{k} = \Phi z_{k} + D u_{k} +w_{k},\quad z_0 = \bar{z}_{0}.
 \end{equation}	
 
 \begin{assumption}[Available data]\label{ass:data}
  While the system matrices  $\Phi \in \R^{\dimy \times \dimz}$ and $D \in \R^{\dimy\times \dimu}$ are considered to be unknown, throughout this paper, inputs and outputs, i.e., $u_k$ and $y_k$ are available at time $k$ through measurements. \End
 \end{assumption}

 To present the main results, we initially assume that the disturbance realizations $w_k$ are measured. We note the availability of disturbance measurements is common in the context of classic disturbance-feedback control schemes \cite{Oldewurtel2008}. It is also relevant for application domains such as energy systems or building control \cite{Drgona2020}. 
  For the cases of unmeasured disturbances, we rely on the estimation methods from~\cite{Pan21s}  and we conduct a robustness analysis of the proposed scheme concerning estimation errors.

 Instead of having exact knowledge about the distributions of the initial condition $\bar{Z}_{0}$ and of  the disturbances  $W_{k}$, $k \in \N $, we assume the following.
  \begin{assumption}[Disturbance with given first two moments]\label{ass:knownmoments} 
    For all 	$ k \in \N$, we consider identically independently distributed (i.i.d.) disturbances with zero mean
       $ \mean[W_k] =0$ and covariance  $\covar[W_{k}]= \Sigma_W$. The uncertain initial condition $\bar{Z}_{0}$, with mean $\mean[\bar{Z}_{0}]$ and covariance $\Sigma[\bar{Z}_{0}]$, is stochastically independent of the disturbances $W_k, k \in \N$.
     \End
     \end{assumption}
 In practice, the knowledge of the first two moments—mean and covariance—of the disturbance can be obtained through empirical estimation with the previous measured or estimated disturbances. 

Let the set of all random variables from $\splx{\dimv}$ with mean $\mean_V \in \R^\dimv$ and covariance $\Sigma_V \in \R^{\dimv \times \dimv}$ be 
\begin{equation}\label{eq:ambiguity_set}
\mcl{D}(\mean_V,\Sigma_V) \doteq \{V \,|\, \mean[V] = \mean_V, \, \Sigma[V] =  \Sigma_V \}.
\end{equation} 
 Assumption~\ref{ass:knownmoments} implies 
     $W_k \in \mcl{D}(0,\Sigma_W)$. Furthermore, the i.i.d.-ness of $W_k$ implies $W_{[0,N-1]} \in \mcl{W}$ with
    \begin{equation}\label{eq:uncertainty_RV}
    	\mcl W\doteq \left\{	W_{[0,N-1]}\, \middle|\,
    	\begin{gathered}
    		\forall i \neq k, \quad i,k \in [0, N-1],  \\
    		W_k  \in \mcl{D}(0,\Sigma_W ),\,\Sigma[W_k,W_i] = 0
    	\end{gathered}  \right\}.% \subset  \splx{\dimw}. 
    \end{equation}
We present our main results assuming that the disturbances are i.i.d. with zero mean. We comment on the extensions to non-i.i.d. settings in Remark~\ref{rem:non_iid}. 

One---not necessarily minimal---state-space realization of the ARX model~\eqref{eq:ARX}  with $Z_k$ as its state is given by
\begin{subequations}\label{eq:extended_dyna}
	\begin{align}
		Z\inst{k+1} &= \widetilde{A}  Z\inst{k} + \widetilde{B} U\inst{k}+ \widetilde{E} W_{k},\quad Z_0 =z_{\ini} \label{eq:RVdynamics_u}
	\end{align}
	with
$
	\widetilde{A} = \begin{bmatrix}
		\bar{A}\\
		\Phi
	\end{bmatrix}, \quad  	 \widetilde{B} = \begin{bmatrix}
		\bar{B}\\
		D
	\end{bmatrix}, \quad  	 \widetilde{E} = \begin{bmatrix}
		0\\
		I_{\dimy}
	\end{bmatrix}.
$ \vspace*{6pt}
  Given the structure of $Z_k$ in \eqref{eq:extendedstate}, the extended states $Z_{k+1}$ and $Z_k$ share overlapping components $	\tra{U}_{[k-\lag+1,k-1]}$ and 	$\tra{Y}_{[k-\lag+1,k-1]}$.
  Hence, we note that $\bar{A}$ and $\bar{B}$ are of known structure and do not depend on the ARX system matrices $\Phi$ and $D$, that is
	\[
	\bar{A} = \left[  \scalemath{0.75}{
		\begin{array}{cccc}
			0 & I_{(\Tini-1)\dimu} & 0&0 \\
			0_{\dimu\times\dimu} 	&0 & 0_{\dimu\times\dimy}& 0\\ 
			0	& 0& 0& I_{(\Tini-1)\dimy} 
	\end{array}}
	\right], \quad \bar{B} =\left[  \scalemath{0.75}{
		\begin{array}{cccc}
			0  \\
			I_{\dimu} \\ 
			0	
	\end{array}}
	\right].
	\]
	Moreover, with $Y_k$ as the last component of $Z_{k+1}$, we have
	\begin{equation}\label{eq:extend_dyna_output}
		Y_k = \widetilde{E}^\top Z_{k+1}.
	\end{equation}
\end{subequations}

As shown in \cite{DePersis19,Berberich2021t,Sadamoto2022}, the benefit of considering \eqref{eq:extended_dyna} is that we may leverage data-driven state-feedback design concepts to obtain output feedback controllers. 
To this end,	we make the following assumption.
\begin{assumption} 
The pair $(\widetilde{A},\widetilde{B})$ from  \eqref{eq:extended_dyna}  is stabilizable. \End
\end{assumption}
This assumption ensures the stabilizability of the system \eqref{eq:extended_dyna} for $W=0$.
In the presence of stochastic disturbances, we have the following result.
	\begin{lemma} \label{lem:controllable}  The pair
		$(\widetilde{A},\left[\widetilde{B},\widetilde{E}\right])$ from \eqref{eq:extended_dyna}  is  controllable regardless of the ARX system matrices $\Phi$ and $D$. \End
	\end{lemma}
	See Appendix~A for the detailed proof.
\begin{remark}[Disturbances and measurement noise]
	Note that the ARX model~\eqref{eq:ARX} inherently incorporates the process disturbance $W_k$ but it does not explicitly consider measurement noise.
To include additive measurement noise $M_k \in \splx{\dimy}$, one can modify the state-space realization~\eqref{eq:extended_dyna} with the output channel~\eqref{eq:extend_dyna_output} replaced by $\hat Y_k = \widetilde{E}^\top Z_{k+1} +M_k$. The distinction between $\hat{Y}_k$ and $Y_k$ is important since the output $Y_k$ contributes directly to the system dynamics as it is part of $Z_{k+1}$, while  $M_k$ acts on the feedback channel. 
While in Section~V-D we consider the estimation of the disturbance, the detailed robustness analysis of data-driven stochastic control with noisy measurements is postponed to future work. We conjecture that the approaches proposed by \cite{Coulson2019,Berberich20} provide an avenue in this direction. \End
\end{remark}

\section{Stochastic Output Feedback Predictive Control}	
Next, we present a conceptual framework for stochastic output feedback predictive control.  We commence with a model-based exposition and then we discuss the challenges of data-driven settings with guaranteed closed-loop properties. 

At time $k$, consider the first two moments of the initial condition $\bar{Z}_k \in \splx{\dimz}$ and of the elements of disturbance sequence $\seq{W}_k \doteq W_{[k,k+N-1]}$  given as per Assumption~\ref{ass:knownmoments}. 
At time $k$, for a finite prediction horizon $N \in \N^+$, let the predicted random-variable input sequence  be written as	$\seq U_k \doteq [U_{0|k}^\top,U_{1|k}^\top,\cdots,U_{N-1|k}]^\top$ with $U_{t|k}$ as the predicted input at time $k+t$ for $t\in \I_{[0,N-1]}$. Similarly, we define the predicted output $Y_{t|k}$, the predicted extended state $Z_{t|k}$, and their sequences $\seq Y_{k}$, $\seq Z_{k}$, respectively.
	 The conceptual Optimal Control Problem (OCP) to be solved at each time step $k$ reads  
		\begin{subequations}\label{eq:stochasticOCP}
		\begin{gather}
			\min_{\substack{
					\seq{Z}_k, \seq{U}_k, \seq{Y}_k , \\ \seq{K}_k , \seq{\bar{u}}_k
			}} \, \max_{\substack{\bar{Z}_k \in \mcl D_{k}, \\ \seq W_k \in \mcl{W}}}
		  J_N\left(\seq U_k , \seq Y_k , Z_{N|k},z_k\right)   \label{eq:OCP_RV_Obj} \\
			\text{s.t. } 	\forall \bar{Z}_k\in \mcl{D}_k, \forall W_{[k,k+N-1]} \in \mcl{W}, \forall t \in \I_{[0,N-1]}  \nonumber\\
				Z\inst{t+1|k}= \widetilde{A}  Z\inst{t|k} + \widetilde{B} U\inst{t|k}+ \widetilde{E} W_{k+t},\quad Z_{0|k} = \bar{Z}_k\label{eq:OCP_RV_dynamic_Z},\\
			Y\inst{t|k} = \widetilde E^\top Z\inst{t+1|k},\label{eq:OCP_RV_dynamic_Y}\\
		U_{t|k} = \bar{u}_{t|k} + K_{t|k} Z_{0|k}+\textstyle{\sum_{j=0}^{t-1}} K_{t,j|k}W_{k+j},  \label{eq:AffinePolicy}\\
						\prob [a_{u,i}^\top U_{t|k} \leq 1 ]\geq 1 - \varepsilon_u,   \,\forall i \in \I_{[1,N_u]},  \label{eq:chance_U} \\
					\prob [a_{y,i}^\top Y_{t|k}  \leq  1 ]\geq 1 - \varepsilon_y,  \, \forall i \in \I_{[1,N_y]},     \label{eq:chance_Y}\\
				 \mean[Z_{N|k}]	 \in \mathbb{Z}_\text{f}, \quad \mean[\|Z_{N|k}\|_\Gamma^2]\leq \gamma.\label{eq:OCP_RV_terminal}
				\end{gather} 
	\end{subequations}
       As the knowledge of the initial condition and of the disturbance is limited to the first two moments, we consider a distributionally robust formulation where the initial condition and the disturbance sequence are taken from the ambiguity sets $\mcl D_k \doteq \mcl{D}(\mean[\bar{Z}_k],\Sigma[\bar{Z}_k] )$ \eqref{eq:ambiguity_set} and $\mcl W$ \eqref{eq:uncertainty_RV}. 
       Hence, \eqref{eq:stochasticOCP} is a (weakly) distributionally robust OCP.\footnote{We refer to \eqref{eq:stochasticOCP} as  a weakly  distributionally robust OCP, as it does not consider any uncertainty for the moments as such. The interested reader is referred to \cite{Pan2023} for an OCP reformulation (but no closed-loop analysis) considering this stronger formulation.} It calls for a min-max optimization
       of the objective functional $J_N$ in \eqref{eq:OCP_RV_Obj} which is given by      
 	 \begin{align*}
      &J_N\left(\seq{U}_k,\seq Y_k, Z_{N|k},z_k\right) \doteq \\
        &\mean  \left[\sum_{t=0}^{N-1} (\|Y_{t|k}\|_Q^2+\|U_{t|k}\|_R^2) +\|Z_{N|k}\|_P^2 \,\middle|\, Z_{0|k}\relx = z_k \right].        
     \end{align*}
    The objective function is the sum of the conditional expected value of the predicted stage cost of inputs and outputs in random variables with $R=R^\top\succ 0 $ and $Q=Q^\top\succeq 0$. Moreover, the matrix $P=P^\top \succeq 0$ specifies the terminal cost with respect to the predicted state. 
    Rather than considering the expectation for all outcomes $\omega \in \Omega$, we focus on the conditional expectation for the outcomes $\{\omega\,|\,Z_{0|k}\relx = z_k\}$, where  the realization $z_k \in \R^\dimz$ is constructed by stacking past $\Tini$ measured inputs and outputs following the structure of \eqref{eq:extendedstate}.  Similar to \cite{Hewing2020}, this approach allows including the measured $z_k$ to minimize the objective while considering the stochastic $\bar{Z}_k$ to satisfy the constraints. As will be shown in Section~\ref{sec:stability}, the stochastic initial condition $\bar{Z}_k$ can be designed to ensure recursive feasibility, while $z_k$ in \eqref{eq:OCP_RV_Obj} serves the purpose of establishing average performance decay.  
    
    Causal and affine feedback policies are required through \eqref{eq:AffinePolicy}, where the current input depends on the realizations of the initial condition and of the past disturbances \cite{Goulart2006}. Compactly, we define $\seq{\bar u}_k \in \R^{N\dimu}$ and $\seq K_k \in R^{N\dimu \times (\dimz + N\dimw)} $ as
    \begin{subequations}\label{eq:compact_gains}
    	\begin{gather}
    		\seq{\bar{u}}_k = [\bar{u}_{0|k}^\top, \bar{u}_{1|k}^\top,\cdots,\bar{u}_{N-1|k}^\top]^\top,\\
    		\seq K_k  =   \left[\begin{smallmatrix}
    			K_{0|k}&		\mbf{0} &\mbf{0} &\cdots& \mbf{0}\\
    			K_{1|k} &	K_{1,0|k} &\mbf{0} & \cdots &   \mbf{0}\\
    			\vdots  & \vdots 	 & \ddots  & \ddots & \vdots \\
    			K_{N-1|k}&	K_{N-1,0|k}& \cdots  &	K_{N-1,N-2|k}& \mbf{0}\\
    		\end{smallmatrix}\right],
    	\end{gather}
    \end{subequations}
which transforms  \eqref{eq:AffinePolicy} into the following compact form
\[
\seq U_k = \seq{\bar{u}}_k + \seq{K}_k [ Z_{0|k}^\top,\seq{W}_k^\top]^\top.
\]
Additionally, we impose chance constraints on $U_{t|k}$ and $Y_{t|k}$ for all $t \in \I_{[0,N-1]}$  in \eqref{eq:chance_U}--\eqref{eq:chance_Y} as individual half-space constraints. That is, $N_u$ input constraints with individual probability of $1-\varepsilon_u$ are defined, where  $a_{u,i} \in \R^\dimu$ for $i \in \I_{[1,N_u]}$. Similarly, $N_y$ output constraints are included.

    Similar to the design of terminal ingredients in the state-feedback stochastic MPC \cite{Pan2023a,farina13probabilistic}, we specify the terminal constraints \eqref{eq:OCP_RV_terminal}. These constraints require the expected value of $Z_{N|k}$ to be within the set $\Zf \subseteq \R^{\dimz}$, and the terminal covariance weighted by $\Gamma \in \R^{\dimz\times\dimz}$ to be smaller than $\gamma \in \R^+$. 
    Furthermore, we require that the terminal ingredients satisfy the following assumption.
	\begin{assumption}[Terminal ingredients]\label{ass:Lypunov}
		Consider the system matrices $\widetilde{A} \in \R^{\dimz \times \dimz}$, $\widetilde{B} \in \R^{\dimz \times \dimu}$, and $\widetilde{E} \in \R^{\dimz \times \dimw}$ from~\eqref{eq:extended_dyna}.
		There exist matrices $P = P^\top \succeq 0$ $\in \R^{\dimz\times \dimz} $, $\Gamma = \Gamma^\top \succeq 0 \in \R^{\dimz \times \dimz}$, $K \in \R^{\dimu \times \dimz}$, and a positive real number $\gamma\in \R^+$,  such that   $A_K= \widetilde{A}+\widetilde{B}K$ is Schur stable, and
		\begin{subequations}
			\begin{gather}
				A_K^\top  P  A_K - P  = -K^\top R K  - A_K^\top \widetilde{E}Q \widetilde{E}^\top A_K, \label{eq:PK} \\
				A_K^\top  \Gamma  A_K - \Gamma  = -I_{\dimz}, \label{eq:Gamma} \\
				\gamma =  \lambda_{\text{max}}(\Gamma) \cdot\trace\left(\Sigma_W\widetilde{E}^\top \Gamma \widetilde{E}\right).\label{eq:gamma}
			\end{gather}
		\end{subequations}
		Moreover, there exists a set $\Zf\subseteq\R^{\dimz}$ such that for all $Z_{N|k} $  satisfying \eqref{eq:OCP_RV_terminal}, we have
		$	A_K\mean[Z_{N|k}] \in \Zf$,
		$	U_{N|k} = K Z_{N|k}$ satisfies  \eqref{eq:chance_U}, and
		$	Y_{N|k} = \widetilde{E}^\top A_K Z_{N|k} + W_{k+N}$ satisfies \eqref{eq:chance_Y}.
		\End
	\end{assumption}

	Utilizing the not necessarily minimal state-space representation  \eqref{eq:extended_dyna}, Assumption~\ref{ass:Lypunov} modifies the terminal ingredients of the model-based state-feedback case \cite{farina13probabilistic} to the output-feedback case. Specifically, \eqref{eq:PK} is a Lyapunov equality tailored to the objective function \eqref{eq:OCP_RV_Obj} with inputs and outputs; \eqref{eq:Gamma}--\eqref{eq:gamma} ensure the $\Gamma$-weighted norm of the covariance of the terminal extended state to be bounded in the presence of stochastic disturbances. Note that \eqref{eq:PK}--\eqref{eq:Gamma}  are Lyapunov equalities. Thus, the solutions $P$ and $\Gamma$ exist if  $A_K= \widetilde{A}+\widetilde{B}K$ is Schur stable and if $K^\top R K  + A_K^\top \widetilde{E}Q \widetilde{E}^\top A_K$ is positive definite. In Section~\ref{sec:stability}, we illustrate the crucial role that the Lyapunov equalities \eqref{eq:PK}--\eqref{eq:Gamma} play for ensuring closed-loop properties. \vspace*{1mm}
 
	The main challenges of adapting \eqref{eq:stochasticOCP} to a data-driven predictive output-feedback setting are threefold: i) the data-driven tractable reformulation of \eqref{eq:stochasticOCP}  involves infinite-dimensional $\mcl{L}^2$ random variables and chance constraints with distributional ambiguity. We address this issue in Section~\ref{sec:datadriven} using ideas from  \cite{Pan21s,Pan2023}. ii) the design of a predictive control algorithm with recursive feasibility guarantees can be achieved via the design of the initial condition, see~Section~\ref{sec:recursive}.  iii) additionally, the terminal ingredients need to be designed without explicit knowledge of system matrices, cf.~Section~\ref{sec:terminal}.	

	\begin{remark}[Witsenhausen's point of view]
		The seminal counter\-{}example of Hans S. Witsenhausen \cite{Witsenhausen68} shows that in stochastic optimal control problems for linear systems with quadratic costs and Gaussian noise affine output-feedback policies are not necessarily optimal. The crucial aspect in this regard is the information available to the controller at each time step, see \cite{Witsenhausen68,Witsenhausen71} and the more recent treatment in \cite{Mitter99}. 
		Hence it is fair to ask whether the affine policies encoded in \eqref{eq:AffinePolicy} jeopardize optimality? For the unconstrained case, the answer is negative as OCP \eqref{eq:stochasticOCP} is based on what Witsenhausen calls an \textit{information pattern with total recall}~\cite{Witsenhausen71}. Put differently,  the initial condition of OCP~\eqref{eq:stochasticOCP}  characterizes the entire extended state variable \eqref{eq:extended_dyna} which in the absence of inequality constraints avoids the issues of the counterexample.~\End 
	\end{remark}

\section{Data-Driven Uncertainty Propagation\label{sec:datadriven}}

\subsection{Basics of Polynomial Chaos Expansions}

We briefly recall Polynomial Chaos Expansion (PCE) and its application to dynamics in $\mathcal{L}^2$ random variables~\eqref{eq:ARX}. PCE dates back to Norbert Wiener \cite{wiener38homogeneous}, and we refer to \cite{sullivan15introduction} for  introductions and to \cite{paulson14fast,Mesbah14,Ou21} for applications in control.

 Due to the $\mathcal L^2$ nature of the considered probability space, there exists an orthogonal polynomial basis $\{\phi^j(\xi)\}_{j=0}^\infty$ that spans $\spl$.  Using the shorthand notation $\phi^j \relx\doteq \phi^j \left(\xi\relx\right)$, we have
\[
  \langle \phi^i , \phi^j  \rangle \doteq \int_{\Omega} \phi^i(\omega)\phi^j(\omega) \diff \mu(\omega) = \delta^{ij}\norm{\phi^j}^2,
\]
where $\delta^{ij}$ is the Kronecker delta and $\norm{\phi^j}\doteq \sqrt{\langle \phi^j,\phi^j\rangle}$ is the induced $\mathcal L^2$  norm  of $\phi^j \in \spl$. We note that it is customary in PCE to consider $\phi^0= 1$ while all other basis functions $\phi^j, j\geq 1$ have zero mean.

The PCE of a random variable $V\in \spl$ with respect to the basis $\{\phi^j(\xi)\}_{j=0}^\infty$ is
\[
 V = \sum_{j=0}^{\infty}\pcecoe{v}{j} \phi^j(\xi)  \text{ with } \pcecoe{v}{j} = \dfrac{\langle V, \phi^j \rangle}{\norm{\phi^j}^2},\]	
 where $\pcecoe{v}{j} \in \R$ is called the $j$-th PCE coefficient.
Accordingly, by applying PCE component-wise, the $j$-th PCE coefficient of a vector-valued random variable $V\in\splx{n_v}$ reads
$\pce{v}^j = $ $ [ \pcecoe{v}{1,j},  \pcecoe{v}{2,j},  \cdots, \pcecoe{v}{n_v,j} ]^\top$
where  $\pcecoe{v}{i,j}$ is the $j$-th PCE coefficient of the component $V^i$ of $V$. 
For the sake of compact notation, we denote 
$\pcecoe{v}{[0,L-1]} \doteq [\pcecoe{v}{0\top},\pcecoe{v}{1\top},\dots,\pcecoe{v}{L-1\top}]^\top \in \R^{\dimv L}$  as the vectorization of the coefficients over the PCE dimensions and  $\hankel_1(\pcecoe{v}{[0,L-1]}) \doteq [\pcecoe{v}{0},\pcecoe{v}{1},\dots,\pcecoe{v}{L-1}] \in \R^{\dimv \times L}$  as their horizontally stacked matrix, cf. Definition~\ref{def:Hankel}.

\begin{definition}[Exact PCE representation \cite{muehlpfordt18comments}]
    The PCE of the  random variable $V\in \splx{n_v}$ is said to be exact   with dimension $L \in \mathbb{N}^+$ in the basis $\{\phi^j(\xi)\}_{j=0}^\infty$  if $	V - \sum_{j=0}^{L-1} \pcecoe{v}{j}\phi^j(\xi)=0.$ \End
\end{definition}
 We  remark that with an exact PCE of  finite dimension, the expected value and covariance of $V\in \splx{n_v}$ can be efficiently calculated from the coefficients 
\begin{equation}\label{eq:PCEmoments}
	\mean\big[V\big] = \pce{v}^0 , \,   \covar \big[V\big] ={ \sum_{j=1}^{L-1} } \pce{v}^j\pce{v}^{j\top}\norm{\phi^j}^2.
\end{equation}

Specifically, using the first two moments of $V\in \splx{n_v}$, we can construct a finite-dimensional conceptual basis that allows an exact PCE of $V$. To this end, we consider the polynomials in $\xi \in \splx{n_\xi}$ with degree of at most $1$  
\begin{equation}\label{eq:P1}
 \psi^0(\xi)=1, \psi^j(\xi) = \xi^j, j\in 
 \I_{[1,n_\xi]},
\end{equation}
where $\xi^j \in \spl$ is the $j$-th component of $\xi$.
 
\begin{lemma}[Exact PCEs by first two moments] \label{lem:finite_PCE}
   	Consider $V \in\splx{\dimv}$ with mean $\mean[V] \in \R^\dimv$ and covariance $\Sigma[V] \succeq 0$. For all $ \pce M  \in \R^{\dimv \times n_\xi}$ satisfying
   	$\pce M \pce M^{\top} =\Sigma[V],$ there exists  $\xi \in \splx{n_\xi}$ with $\mean[\xi]=0$ and $\Sigma[\xi] = I_{n_\xi}$ such that $ V= \mean[V] + \pce M \xi$  holds.  	
   	Using the basis  $\{\psi^j(\xi)\}_{j=0}^{n_\xi}$ with $\psi$ from \eqref{eq:P1}, an exact and finite-dimensional PCE of $V$ is obtained as
   	\begin{equation}\label{eq:PCEGaussian}
   	V=\mean[V]+	\pce M  \xi= \sum_{j=0}^{n_\xi} \pcecoe{v}{j} \psi^j(\xi)  
   	\end{equation}
   	with $\pcecoe{v}{0} = \mean[V]$, $\hankel_1(\pcecoe{v}{[1,n_\xi]}) = \pce M $.
  \End
   	\end{lemma}
See Appendix~B for the detailed proof. 
    
Notice that in \eqref{eq:PCEGaussian} the coefficients encode the information of the first two moments, while the stochastic germ $\xi$ preserves the information of the distribution. In other words, the stochastic germ $\xi$ serves as a normalized version of $V$. Common choices for the decomposition $\pce	M \pce M^{\top} =\Sigma[V]$ include the principal square root and the Cholesky decomposition. 
    
To exactly represent random variables other than $V$, an alternative approach is to construct an orthogonal basis $\{\phi^j(\xi)\}_{j=0}^{\infty}$ that spans $\spl$ using the Gram-Schmidt procedure. This method involves higher-degree polynomials of $\xi$~\cite{Witteveen2006}. However, in the following, by including all uncertainties of \eqref{eq:stochasticOCP} into the stochastic germ $\xi$, we construct a finite-dimensional basis consisting of polynomials with degrees of at most $1$. This basis, as detailed in the next section, enables exact PCE descriptions of all random variables involved in \eqref{eq:stochasticOCP}. 

\subsection{Data-Driven Uncertainty Propagation via PCE }\label{sec:data_driven_UQ}
     Given the  first two moments of $\bar{Z}_k$, $W_{k+t}$ (as per Assumption \ref{ass:knownmoments})
    and the covariance decompositions  $ \pce M_{k} \pce M_{k}^\top = \Sigma[\bar{Z}_k]$ and  $ \pce M_{w} \pce M_{w}^\top = \Sigma_W$, with $\pce M_{k} \in \R^{\dimz \times n_\xi}$ and $\pce M_{w} \in \R^{\dimw \times n_\eta}$, we construct their bases and PCEs via Lemma~\ref{lem:finite_PCE} as  
    \begin{align*}
        \{\psi^j(\xi_{k})\}_{j=0}^{n_\xi}, &\quad  \bar{Z}_{k}= \mean[\bar{Z}_k]+ \pce M_{k}\xi_k, \\  \{\psi^j(\eta_{k+t})\}_{j=0}^{n_\eta},&\quad W_{k+t}= \pce M_w\eta_{k+t} , t \in \I_{[0,N-1]}.
    \end{align*}
     Moreover, the  propagation of uncertainties through \eqref{eq:OCP_RV_dynamic_Z}--\eqref{eq:OCP_RV_dynamic_Y} with  affine policies  \eqref{eq:AffinePolicy} can be achieved with exact PCEs in the following orthonormal basis 
    \begin{equation}\label{eq:common_basis}
    	\{\psi^j(\boldsymbol \xi_k)\}_{j=0}^{L-1}, \quad\boldsymbol{\xi}_k =  [\xi_k^\top,\eta_{k}^\top,...,\eta_{k+N-1}^\top]^\top,
    \end{equation}
    where $L = Nn_\eta+n_\xi+1$, cf. \cite[Prop. 1]{Pan21s}. Due to the weakly distributionally robust formulation of OCP~\eqref{eq:stochasticOCP}, we do not require the exact information of $\xi_k$ and $\eta_k$ for  OCP~\eqref{eq:stochasticOCP}. Rather we consider all possible  $\xi_k \in \mcl D (0, I_{n_\xi})$ and $\eta_{k+i} \in \mcl  D (0, I_{n_\eta}) $, $i \in \I_{[0,N-1]}$, cf. \eqref{eq:ambiguity_set}. In other words, the constraints of OCP~\eqref{eq:stochasticOCP} are required to hold  for all $\boldsymbol \xi_k \in \mcl D (0, I_{N n_\eta+ n_\xi})$.
    
    Applying Galerkin projection~\cite{GhanSpan03} onto the basis~\eqref{eq:common_basis}  yields the dynamics of the PCE coefficients, for all $ j\in \I_{[0,L-1]}$,
    \begin{subequations}\label{eq:ARX_PCE}
    	\begin{align}
    		\pce{z}^j_{t+1|k} &= \widetilde{A} \pce{z}^j_{t|k}+  \widetilde{B}  \pce{u}^j_{t|k}   +  \widetilde{E} \pce{w}^j_{k+t}, \quad   \pce{z}^j_{0|k} = \bar{\pce{z}}^j_{k},\, \\
    		\pce{y}^j_{t|k} &= \Phi \pce{z}^j_{t|k}+ D \pce{u}^j_{t|k}   + \pce{w}^j_{k+t}, \quad   \forall t \in \I_{[0,N-1]},
    	\end{align}
    \end{subequations}	
   where the PCE coefficients of $\bar{Z}_k$ and $W_{k+t}$, $t \in \I_{[0,N-1]}$ are
  \begin{equation}\label{eq:coefficients_z_w}
    \begin{gathered}
    \pcecoe{\bar z}{0}_k = \mean[\bar{\pce{Z}}_k], \, \hankel_1(\bar{\pce{z}}^{[1,n_\xi]}_{k})= \pce M_{k}, \, \pcecoe{\bar z}{[n_\xi+1, L-1]}_k = 0, \\
  \hankel_1(\pce{w}^{\mcl I(t)}_{k+t})= \pce M_w,\, \pcecoe{w}{j}_{k+t}=0,\,\forall j \in \I_{[0,L-1]} \backslash \mcl I(t)  ,
    \end{gathered}
    \end{equation}
    where the range $\mcl I(t)= [tn_\eta+n_\xi+1,(t+1)n_\eta+n_\xi]$ corresponds to the PCE indices of the basis functions related to the disturbance $W_{k+t}$ in the basis \eqref{eq:common_basis}.

    Note that the distributional ambiguity surrounding $W_{k+t}$ and $\bar{Z}_k$ leads to a situation where the predictions obtained through \eqref{eq:OCP_RV_dynamic_Z}--\eqref{eq:OCP_RV_dynamic_Y}  are characterized by ambiguous distributions. However, representing all random variables by their PCEs in the basis \eqref{eq:common_basis}, we obtain deterministic coefficients without uncertainty which correspond to the mean and the covariance decomposition. Put differently, the distributional ambiguities are encoded in the basis $\{\psi^j(\boldsymbol \xi_k)\}_{j=0}^{L-1}$ with $ \boldsymbol{\xi}_k \in \mcl{D}(0,I_{Nn_\eta+n_\xi})$, cf. \eqref{eq:ambiguity_set}. This crucial observation allows to reformulate the distributional robust chance constraints to tractable second-order cone constraints. 
    To this end, we recall a result based on \cite[Thm.~3.1]{Calafiore2006}, which is a generalization of the Chebyshev-Cantelli inequality. This result also appeared previously as \cite[Prop.~1]{Pan2023}.
    \begin{lemma}[PCE for chance constraints \cite{Pan2023}]\label{lem:chance_reformulation}
    	Consider a  random variable $V \in \splx{\dimv}$ with its PCE $V(\boldsymbol{\xi}_k) = \sum_{j=0}^{L-1} \pcecoe{v}{j} \psi^j(\boldsymbol{\xi}_k)= \pcecoe{v}{0}+\hankel_1(\pcecoe{v}{[1,L-1]})\boldsymbol{\xi}_k$ regarding the basis~\eqref{eq:common_basis}. For $a \in \R^{\dimv}$, the distributionally robust chance constraint
    	\begin{subequations}
    		\begin{equation}\label{eq:chance_constraints}
    	       \prob [a^\top V(\boldsymbol{\xi}_k) \leq 1 ]\geq 1 - \varepsilon,\, \forall \boldsymbol{\xi}_k \in \mcl{D}(0,I_{Nn_\eta+n_\xi})	
    	    \end{equation}
        is equivalent to
            \begin{equation}\label{eq:chance_reformulation}
    	       a^\top\pcecoe{v}{0} +	  \sigma (\varepsilon) \|a^\top \hankel_1(\pcecoe{v}{[1,L-1]} ) \|  \leq 1
            \end{equation}
        with  $ \sigma (\varepsilon)=\sqrt{ (1-\varepsilon)/\varepsilon }. $ \End
    		\end{subequations}
    	\end{lemma}  
    Without distributional ambiguity and if $V(\boldsymbol{\xi}_k)$ is known to be Gaussian distributed,  one may replace  $\boldsymbol{\xi}_k \in \mcl{D}(0,I_{Nn_\eta+n_\xi})$ by $\boldsymbol{\xi}_k \sim \mcl{N}(0,I_{Nn_\eta+n_\xi})$. Then \eqref{eq:chance_constraints} is equivalent to \eqref{eq:chance_reformulation}  if $\sigma(\varepsilon)$ is chosen according to the standard normal table \cite{Calafiore2006}. 
  
As the system matrices of \eqref{eq:ARX} are unknown, we recall the data-driven representation of stochastic LTI systems proposed in \cite{Pan21s}, which relies on a stochastic variant of Willems' fundamental lemma~\cite[Thm.~1]{Willems2005}.

\begin{definition}[Persistency of excitation \cite{Willems2005}] Let $T, N \in \set{N}^+$. A sequence of inputs $\trar{u}{T-1}$ is said to be persistently exciting of order $N$ if the Hankel matrix 	$\hankel_N(\trar{u}{T-1})$	is of full row rank.  \End
\end{definition}

\begin{lemma}[Stochastic fundamental lemma~\cite{Pan21s,Faulwasser2022}]\label{lem:RVfundamental} 
	Consider system \eqref{eq:ARX}, its state-space realization  \eqref{eq:extended_dyna}, and its  trajectory of $\mcl L^2(\Omega, \mcl F,\mu;\R^{\dimv})$, 
	$\dimv \in\{ \dimu, \dimw, \dimy\}$
	random variables and the corresponding realization trajectory from \eqref{eq:ARX_realization}, which are $\trar{(U,W,Y)}{T-1}$, respectively, $\trar{(u,w,y)}{T-1}$.  
	Let $\trar{(u,w)}{T-1}$ be persistently exciting of order $\dimz +N$.  
	
\vspace*{0.1cm} Then,
	$ (\widetilde{U}, \widetilde{W},\widetilde{Y})_{[0,N-1]}$ is a trajectory of \eqref{eq:ARX} if and only if there exists $G \in \splx{T-N+1} $ such that 
	\begin{equation} \label{eq:RVfunda}
		\hankel_N(\trar{v}{T-1}) G=\widetilde{V}_{[0,N-1]}
	\end{equation} 
	holds for all $(v, \widetilde{V})\in \{(u,\widetilde{U}), (w, \widetilde{W}),(y,\widetilde{Y})\} $. 
		
  Moreover, for all $j \in \I_{[0,L-1]}$, $\trar{ ( \tilde{\pce u}, \tilde{\pce w},\tilde{\pce y})}{N-1}^j $ is a PCE coefficient trajectory of \eqref{eq:ARX_PCE}
		if and only if there exists  $\pcecoe{g}{j}\in \R^{T-N+1}$ such that 
		\begin{equation} \label{eq:mixed_funda}
			\hankel_N(\trar{v}{T-1}) \pcecoe{g}{j}= \trar{ \tilde{\pce v}}{N-1}^j
		\end{equation}
		holds for all $(v, \tilde{\pce{v}})\in \{ (u,\tilde{\pce{u}}), (w,\tilde{\pce{w}}),(y,\tilde{\pce{y}})\} $.
		\End
\end{lemma}
	
The key insight of the stochastic fundamental lemma is that one can represent the trajectories of $\mcl L^2$  random variables directly in the subspace spanned by their realizations trajectories. Underlying this is the observation that the dynamics of random variables~\eqref{eq:ARX}, its realization dynamics~\eqref{eq:ARX_realization}, and the dynamics of PCE coefficients~\eqref{eq:ARX_PCE} share the same system matrices. For more details, we refer to~\cite{Faulwasser2022}. Moreover,  the tailored stochastic fundamental lemma, i.e. Lemma~\ref{lem:RVfundamental}, does not hinge on the controllability of the pair $(\widetilde{A},\widetilde{B})$ thanks to Lemma~\ref{lem:controllable}.

\subsection{Data-Driven Stochastic Optimal Control}\label{sec:OCP}
    Before arriving at the data-driven reformulation of OCP~\eqref{eq:stochasticOCP}, we derive the PCE reformulation of the objective $J_N$ in \eqref{eq:OCP_RV_Obj}.  
    Recall that \eqref{eq:OCP_RV_Obj} is a conditional expectation for the outcomes $\{\omega\,|\,Z_{0|k}\relx = z_k\}$. 
Combined with  $Z_{0|k} = \bar{Z}_k$ in \eqref{eq:OCP_RV_dynamic_Z} and PCE coefficients of $\bar{Z}_k$ in \eqref{eq:coefficients_z_w}, we obtain the reformulation of the set of corresponding outcomes 
    \begin{equation}\label{eq:initial_con_map}
       \left\{ \omega \,\middle|\,  
        \bar{\pce{z}}_k^0 +  \hankel_1\left(\pcecoe{\bar z}{[1,n_\xi]}_k\right) \xi_k\relx =  z_k\right\}.
    \end{equation}
    At this point, one has two principal options: determine  $\xi_k\relx$ prior to the optimization  or consider it as an extra decision variable. For the further developments, we suppose temporarily that $\xi_k\relx$ is known, while we show how it can be determined in Section~\ref{ssec:Tk}.
  	Given the realization $\xi_k\relx$,
the reformulation of \eqref{eq:OCP_RV_Obj} reads
\begin{align}
 			J_N& = \textstyle\sum_{t=0}^{N-1} \left(\ell_{u, t|k} + \ell_{y, t|k} \right) +\ell_{z, N|k} , \label{eq:obj_reformulation}\\
	 		\ell_{u, t|k}  & \textstyle=  \mean\left[ \|\sum_{j=0}^{L-1}\pcecoe{u}{j}_{t|k}\psi^j(\boldsymbol{\xi}_k) \|_R^2 \,  \middle |\, \xi_k\relx     \right]  \nonumber\\
	 & =  \textstyle\|\pcecoe{u}{0}_{t|k} + \hankel_1(\pcecoe{u}{[1,n_\xi]}_{t|k})\xi_k\relx\|_R^2 +  \sum_{j=n_\xi+1}^{L-1} \|\pcecoe{u}{j}_{t|k}\|^2_R, \nonumber\\
 		\ell_{y, t|k}&\textstyle =  \|\pcecoe{y}{0}_{t|k} + \hankel_1(\pcecoe{y}{[1,n_\xi]}_{t|k})\xi_k\relx\|_Q^2 +  \sum_{j=n_\xi+1}^{L-1} \|\pcecoe{y}{j}_{t|k}\|^2_Q, \nonumber\\
   		\ell_{z, N|k} & \textstyle=  \|\pcecoe{z}{0}_{N|k} + \hankel_1(\pcecoe{z}{[1,n_\xi]}_{N|k})\xi_k\relx\|_P^2 +  \sum_{j=n_\xi+1}^{L-1} \|\pcecoe{z}{j}_{N|k}\|^2_P. \nonumber
 	\end{align}

   Now, we are ready to state the data-driven reformulation of OCP~\eqref{eq:stochasticOCP}.
     At time instant $k$, given the PCE coefficients $\bar{\pce{z}}^{[0,L-1]}_k$  and  $\pce{w}_{k+t}^{[0,L-1]}$, $t\in\I_{[0,N-1]}$ in \eqref{eq:coefficients_z_w}, and the realization $\xi_k\relx$,
    	we solve 
    	\begin{subequations}\label{eq:PCEOCP}
    		\begin{gather}
    			  \min_{\substack{
    			  		\seq{\bar{u}}_k,\seq K_k,
    					\seq{\pce{u}}_k,\seq{\pce y}_k,\seq{\pce z}_k,\seq{\pce g}
    				}
    			} \, 
    		J_N\left( \seq{\pce{u}}_k, \seq {\pce{y}}_k,\pcecoe{z}{[0,L-1]}_{N|k},\xi_k\relx\right)  \\
    			\text{s.t. }	 \forall  j\in \set{I}_{[0,L-1]},\, \forall  t\in \set{I}_{[0,N-1]}\nonumber   \\
    			\quad 
    			\left[
    			\begin{array}{l}
    				\hankel_{N+\Tini}(\trar{u}{T-1})\\
    				\hankel_{N+\Tini}(\trar{y}{T-1})\\
    				\hankel_{N}(\tra{w}_{[\Tini,T-1]})\\
    			\end{array}
    			\right]
    			\pcecoe{g}{j}	=
    			\left[
    			\begin{array}{l}
    				\pce{u}^j_{[-\Tini,N-1]|k}\\
    				\pce{y}^j_{[-\Tini,N-1]|k}\\     
    				\pce{w}^j_{[k,k+N-1]}\\
    			\end{array}
    			\right]
    			 \label{eq:H_PCE_SOCP_hankel}	\\				
    			\begin{bmatrix}
    				\pcecoe{u}{j}_{[t-\Tini,t-1]|k}\\
    				\pcecoe{y}{j}_{[t-\Tini,t-1]|k} 
    			\end{bmatrix} 	= \pce{z}^j_{t|k},\quad	\pce{z}^j_{0|k} =\bar{\pce{z}}^j_{k} \label{eq:PCEOCP_Ini}\\
   \pcecoe{u}{j}_{t|k}= \delta^{0j}	\bar{u}_{t|k} + K_{t|k} \pcecoe{z}{j}_{0|k}+\textstyle{\sum_{j=0}^{t-1}} K_{t,j|k}\pcecoe{w}{j}_{k+j} \label{eq:PCEOCP_u}\\
      		a_{u,i}^\top\pcecoe{u}{0}_{t|k} +\sigma(\varepsilon_u)	\| a_{u,i}^\top \hankel_1(\pcecoe{u}{[1,L-1]}_{t|k})\|  \leq 1 , \forall i \in \I_{[1,N_u]} \label{eq:PCEOCP_chanceu} \\
    a_{y,i}^\top \pcecoe{y}{0}_{t|k} +	  \sigma(\varepsilon_y)
    \| a_{y,i}^\top \hankel_1 ( \pcecoe{y}{[1,L-1]}_{t|k})\|  \leq 1 , \forall i \in \I_{[1,N_y]} \label{eq:PCEOCP_chancey}\\
    			\pcecoe{z}{0}_{N|k}
    	 \in \Zf, \quad  \textstyle{\sum_{j=1}^{L-1}}\|		\pcecoe{z}{j}_{N|k}\|^2_\Gamma \leq \gamma
    			\label{eq:terminal}
    		\end{gather} 
    	\end{subequations}
    with  $ \sigma (\varepsilon)=\sqrt{ (1-\varepsilon)/\varepsilon } $. Here, $\seq{\pce u}_k $  denotes the shorthand for the collection of all PCE coefficients of $U_{t|k}$, i.e. $\pcecoe{u}{j}_{t|k}$ for $j \in \I_{[0, L-1]}$ and $t \in \I_{[0,N-1]}$. Similarly, we define $\seq{\pce y}_k$ and $\seq{\pce g}$.
    
    If $\Zf \subseteq \R^{\dimz}$ is chosen as a sublevel set of $\|\pcecoe{z}{0}_{N|k}\|_P^2$ with $P \succeq 0$, the data-driven OCP \eqref{eq:PCEOCP} is a second-order cone program. 
    Using interior-point methods to solve \eqref{eq:PCEOCP} typically requires at most $O(\sqrt{n_c})$ iterations with complexity of $O(n_c n_d^2)$ per iteration \cite{Lobo1998}. Here,  $n_c = N(N_u+N_y)+2$ is the number of second-order cone constraints and  $n_d$ denotes the number of decision variables in \eqref{eq:PCEOCP}.
    
    The equivalence of OCP~\eqref{eq:stochasticOCP} and OCP~\eqref{eq:PCEOCP} relies on several factors \cite{Pan2023}. First, the PCE representations of all random variables are exact by the construction of basis~\eqref{eq:common_basis}.
    Second, by Lemma~\ref{lem:RVfundamental}, the data-driven reformulation of the system dynamics of \eqref{eq:OCP_RV_dynamic_Z}--\eqref{eq:OCP_RV_dynamic_Y} to \eqref{eq:H_PCE_SOCP_hankel}--\eqref{eq:PCEOCP_Ini} is exact if the disturbance realization $w_{[\Tini,T-1]}$ is available.
    Third,  the reformulation of the distributionally robust chance constraints from \eqref{eq:chance_U}--\eqref{eq:chance_Y} to \eqref{eq:PCEOCP_chanceu}--\eqref{eq:PCEOCP_chancey} is also exact by Lemma~\ref{lem:chance_reformulation}.  Remarkably,  \eqref{eq:PCEOCP} does not explicitly dependent on the basis $\{\psi^j(\boldsymbol \xi)\}_{j=0}^{L-1}$ and thus  the knowledge of the exact distribution of $\boldsymbol{\xi} \in \mcl{D}(0,I_{Nn_\eta+n_\xi})$ is not required. In addition,  combined with the PCE coefficients $\bar{\pce{z}}^{[0,L-1]}_k$  and  $\pce{w}_{k+t}^{[0,L-1]}$, $t\in\I_{[0,N-1]}$ in \eqref{eq:coefficients_z_w}, the disturbance feedback policies  \eqref{eq:PCEOCP_u} further imply 
    \begin{equation}\label{eq:causality}
    	\pcecoe{u}{j'}_{t|k} = 0,\quad \forall j' \in \I_{[n_\xi+tn_\eta+1,L-1]}.
    \end{equation}
\section{Data-Driven Stochastic Output Feedback } \label{sec:MPC}

\subsection{Recursive Feasibility with Backup Initial Condition}\label{sec:recursive}
Next, we present our output-feedback stochastic data-driven predictive control scheme based on OCP~\eqref{eq:PCEOCP}. With the measured past $\Tini$ input and output realizations, i.e. the current realization $z_k$ of the extended state $Z_k$,  one can set the initial condition of \eqref{eq:PCEOCP} as
\begin{equation} \label{eq:consistency_prime}
	\bar{\pce{z}}^j_{k} = 	\delta^{0j} z_k, j \in \I_{[0,L-1]}
\end{equation}
with $	\delta^{0j}$ as the Kronecker delta.  In other words, we construct a deterministic initial condition by setting the measured value as the 0-th PCE coefficients, while setting higher-order PCE coefficients that represent stochastic uncertainties to zero. Henceforth, we refer to \eqref{eq:consistency_prime} as the \textit{measured initial condition}.

However, at time step $k$, OCP~\eqref{eq:PCEOCP} can be infeasible, e.g., for disturbances $W$ with infinite support if at step $k-1$ a large realization $w_{k-1}$ occurs. If so, we rely on a so-called \textit{backup initial condition} to ensure recursive feasibility. For the underlying concept from model-based stochastic predictive control, we refer to \cite{Farina2015,Hewing18s}.

Similar to deterministic predictive control, one way for constructing a backup initial condition $\bar{Z}_{k}$ is to consider its predicted value based on the optimal solution of the last OCP solved, i.e. $\Zkini = Z^{\star}_{1|k-1}$. However, the one-step prediction of $Z^{\star}_{1|k-1}$ by \eqref{eq:extended_dyna} with the last initial condition $\bar{Z}_{k-1}$ involves the stochastic uncertainty of $W_{k-1}$. This leads to growth of the PCE basis,  e.g.,  see \cite{Pan2023a}. In this paper, using the insights of Lemma~\ref{lem:finite_PCE}, we update the basis \eqref{eq:common_basis} at each time instant to keep its dimensionality constant. As shown in the next result, for all $k\in \N$, we can state a feasibility-preserving initial condition for OCP~\eqref{eq:PCEOCP} in this updated basis.

At time instant $k-1$,  suppose OCP~\eqref{eq:PCEOCP} is feasible with the initial condition $\bar{ \pce{z}}^{[0,L-1]}_{k-1}$  with $\bar{ \pce{z}}^{j}_{k-1} =0$ for all $j \in \I_{[1+n_\xi,L-1]}$. The optimal solution of the OCP determines a prediction 
\[Z^{\star}_{1|k-1} =  \pce{z}^{0,\star}_{1|k-1} + \hankel_1\left(\pce{z}^{[1,n_\xi+n_\eta],\star}_{1|k-1}\right)  \begin{bmatrix} \xi_{k-1} \\ \eta_{k-1}
\end{bmatrix}. \]
Observe that the above is an exact PCE of $Z^{\star}_{1|k-1} $ with dimensionality $n_\xi+n_\eta+1$. To  avoid any dimension growth of the PCE basis, we design an orthonormal projection matrix $T_k \in \mathbb{R}^{(n_\xi+n_\eta) \times n_\xi}$, i.e., $T_k^\top T_k = I_{n_\xi} $. 

\begin{proposition}[Recursive feasibility with updated basis]\label{pro:recur}
 Let Assumptions~\ref{ass:data}--\ref{ass:Lypunov} hold. 
 At time instant $k-1$, 
 suppose OCP~\eqref{eq:PCEOCP} is feasible in the basis $\{\psi^j(\boldsymbol{\xi}_{k-1})\}_{j=0}^{L-1}$ from \eqref{eq:common_basis}.
At time instant $k$,  for any  orthonormal matrix $T_k \in \mathbb{R}^{(n_\xi+n_\eta) \times n_\xi}$ satisfying $T_k^\top T_k = I_{n_\xi} $, considering
 $\{\psi^j(\boldsymbol \xi_k)\}_{j=0}^{L-1}$ with 
    \begin{equation}\label{eq:update_xik}
        \xi_k = T_k^\top \begin{bmatrix} \xi_{k-1} \\ \eta_{k-1}
    \end{bmatrix},\, \boldsymbol{\xi}_k =  [\xi_k^\top,\eta_{k}^\top,...,\eta_{k+N-1}^\top]^\top,
    \end{equation}
OCP~\eqref{eq:PCEOCP} is feasible  with the initial condition 	$\bar{\pce{z}}^{[0,L-1]}_{k}$   with $\bar{ \pce{z}}^{j}_{k} =0$ for all $j \in \I_{[1+n_\xi,L-1]}$ and
     	\begin{equation}	\label{eq:moment_matching}
\bar{\pce{z}}^{0}_{k} = \pce{z}^{0,\star}_{1|k-1},\, 	\hankel_1\left(\bar{\pce{z}}^{[1,n_\xi]}_{k}\right) =	\hankel_1\left(\pce{z}^{[1,n_\xi+n_\eta],\star}_{1|k-1}\right) T_k.
   	\end{equation}
 \End
\end{proposition}
The detailed proof is given in Appendix~C. Notice that the updated  $\boldsymbol{\xi}_k$ still satisfies $\boldsymbol{\xi}_k \in \mcl{D}(0,I_{Nn_\eta+n_\xi})$.

\subsubsection{Online Selection of Initial Conditions}
To distinguish the initial conditions, we refer to  \eqref{eq:consistency_prime} as the \textit{measured initial condition} and to \eqref{eq:moment_matching} as the \textit{backup initial condition}. Moreover, we denote the function value of the optimal value function of OCP~\eqref{eq:PCEOCP}  obtained with the measured initial condition \eqref{eq:consistency_prime}  as $V^{\text{m}}_N$  and to its counterpart obtained with the backup initial condition \eqref{eq:moment_matching} as $V^{\text{b}}_N$.
The notion \textit{backup initial condition} refers to the underlying idea of using the measured data as much as possible since it provides feedback. Eventually, we consider an initial condition selection strategy similar to \cite{farina13probabilistic}:
\begin{enumerate}
	\item[i)] At each time instant $k\in \N$, initialize $V^{\text{b}}_N = +\infty$ and $V^{\text{m}}_N= +\infty$.
	\item[ii)]  If the OCP~\eqref{eq:PCEOCP} with the measured initial condition \eqref{eq:consistency_prime} is infeasible,  then re-solve  with the backup one \eqref{eq:moment_matching}.  The optimal cost $V^{\text{b}}_N$ is updated correspondingly.
	\item[iii)]  Else the OCP~\eqref{eq:PCEOCP} with the measured initial condition \eqref{eq:consistency_prime} is feasible, update the resulting optimal cost $V^{\text m}_N$ and compare it to the performance $\hat{J}_{N,k}$ obtained from a feasible candidate solution, the construction of which is detailed in~\eqref{eq:projection} in Appendix~B. If $V^{\text{m}}_N \leq \hat{J}_{N,k}$ holds, apply the optimal input obtained with the measured initial condition. 
	\item[iv)]  Else, i.e., whenever $V^{\text{m}}_N > \hat{J}_{N,k}$ holds, we  re-solve OCP~\eqref{eq:PCEOCP} with the backup initial condition~\eqref{eq:moment_matching}. Note that due to optimality the resulting $V^{\text{b}}_N$ satisfies  $V^{\text{b}}_N\leq \hat{J}_{N,k}$.
\end{enumerate}
It is easy to see that this procedure ensures that the resulting optimal cost at time instant $k$ satisfies
\begin{equation}\label{eq:costdecay}
	V_{N,k} \doteq \min\left\{V^{\text{m}}_N,V^{\text{b}}_N\right\} \leq \hat{J}_{N,k},\quad k\in \N.
\end{equation}

\subsubsection{Design of the Projection $T_k$}\label{ssec:Tk}
With the backup initial condition \eqref{eq:moment_matching} in PCE coefficients, its random-variable counterpart reads as 
\begin{equation}\label{eq:barZk}
	\begin{aligned}\bar{Z}_{k} & = \pcecoe{\bar{z}}{0}_k + \hankel_1\left(\pcecoe{\bar{z}}{[1,n_\xi]}_{k} \right)\xi_k
	\\ 
	&	=  \pcecoe{z}{0,\star}_{1|k-1}+ \hankel_1 \left( \pcecoe{z}{[1,n_\xi+n_\eta],\star}_{1|k-1}                                                  \right) T_k
	T_k^\top  \begin{bmatrix} \xi_{k-1} \\ \eta_{k-1}\end{bmatrix} .
\end{aligned}
\end{equation}
Observe that the orthonormal matrix $T_k \in \mathbb{R}^{(n_\xi+n_\eta) \times n_\xi}$ satisfies  $T_k^\top T_k = I_{n_\xi} $ but $T_k T_k^\top \neq I_{\dimz}$. Hence, with the projection $T_k$ to avoid the dimension growth of the PCE basis, we do not have the equality $\Zkini = Z^{\star}_{1|k-1}$ in random variables as considered in \cite{Pan2023a}. Instead, we design $T_k$ to ensure that both the realizations of the prediction $Z^{\star}_{1|k-1}$ and the backup initial condition $\bar{Z}_k$ coincide with  the current measurement $z_k$.  This leads to a  conditional equality
\begin{equation}\label{eq:conditional_equality}
	\bar{Z}_k\relx = Z^{\star}_{1|k-1}\relx,\quad \forall \omega \in \{ \omega \,|\, Z^{\star}_{1|k-1}\relx = z_k \}.
\end{equation}

\begin{subequations}\label{eq:projection_Tk}
To enforce \eqref{eq:conditional_equality}, we consider $\xi_k \in \spl$  in \eqref{eq:barZk} as a scalar-valued random variable with realization 
	\begin{equation}\label{eq:xi_k}
			\xi_k\relx= \left\|\begin{bmatrix} \xi_{k-1}\relx \\ \eta_{k-1}\relx\end{bmatrix}\right\| \in \mathbb{R}.
	\end{equation}
Moreover, we use the  projection
	\begin{equation}\label{eq:Tk}
			T_k = \frac{[ \xi_{k-1} \relx , \eta_{k-1}^\top \relx]^\top }{\left\|[ \xi_{k-1} \relx , \eta_{k-1}^\top \relx ]^\top \right\|} \in \mathbb{R}^{(1+n_\eta) \times 1}
	\end{equation}
such that 
	\begin{equation*}
			\xi_k\relx= T_k^\top\begin{bmatrix} \xi_{k-1}\relx \\ \eta_{k-1}\relx\end{bmatrix},\quad  	T_k	T_k^\top  \begin{bmatrix} \xi_{k-1}\relx \\ \eta_{k-1}\relx\end{bmatrix} =  \begin{bmatrix} \xi_{k-1}\relx \\ \eta_{k-1}\relx\end{bmatrix}
	\end{equation*}
hold. Applying the above to \eqref{eq:barZk}, we obtain the conditional equality \eqref{eq:conditional_equality}.
Furthermore, recall that the current measurement $z_k$ is the realization of the prediction $Z^{\star}_{1|k-1}$, i.e.,
\[
Z^{\star}_{1|k-1}\relx =  \pcecoe{z}{0,\star}_{1|k-1}+ \hankel_1\left(\pcecoe{z}{[1,1+n_\eta],\star}_{1|k-1} \right)\begin{bmatrix} \xi_{k-1}\relx \\ \eta_{k-1}\relx\end{bmatrix} = z_k.
\]
With $\hankel_1\left(\pcecoe{z}{[2,1+n_\eta],\star}_{1|k-1} \right)=\widetilde E \pce M_w$ by \eqref{eq:ARX_PCE}--\eqref{eq:coefficients_z_w}, we determine $\eta_{k-1}\relx$ via the least-square solution to the above equality.
\begin{equation}\label{eq:eta_estimation}
\eta_{k-1}\relx =   \pce M_w^{\dagger} \widetilde E \left(z_k- \pcecoe{z}{0,\star}_{1|k-1} -  \pcecoe{ z}{1,\star}_{1|k-1}\xi_{k-1}\relx\right).
\end{equation}
\end{subequations}
\subsection{Proposed Predictive Control Algorithm}\label{sec:Algorithm}
We are ready to present the main result, i.e., a data-driven stochastic output-feedback predictive control scheme with closed-loop guarantees based on OCP~\eqref{eq:PCEOCP}. The scheme consists of an offline data collection phase and an online optimization phase containing a procedure for initial condition selection. The overall scheme is summarized in Algorithm~\ref{alg:datadrivenSMPC}.

In the offline phase, a random input trajectory $u_{[0,T-1]}$ is generated to obtain $y_{[0,T-1]}$ under disturbances $w_{[0,T-1]}$.  We discuss the case of unmeasured disturbance data $w_{[0,T-1]}$ in Section~\ref{sec:estimate_disturbance}.
Moreover, the recorded input, output, and disturbance trajectories are used to determine the terminal ingredients proposed in Section~\ref{sec:terminal} and to construct the Hankel matrices for OCP~\eqref{eq:PCEOCP}.

In the online optimization phase,  we assume that the OCP~\eqref{eq:PCEOCP} is initially feasible with the measured initial condition at $k=0$.
At each time instant $k$, before applying the initial condition selection strategy, we  update $\eta_{k-1}\relx$, $T_k$, $\xi_k\relx$ by  \eqref{eq:projection_Tk} using the measured $z_k$, the predicted $\pcecoe{z}{[0,n_\xi+n_\eta],\star}_{1|k-1}$, and the last $\xi_{k-1}\relx$. Subsequently, we update $\hat J_{N,k}$ by evaluating \eqref{eq:obj_reformulation} with $\xi_k\relx$ and the feasible candidate~\eqref{eq:projection}.
After solving OCP~\eqref{eq:PCEOCP} with initial condition selection in Lines~\ref{step:selectBegin}--\ref{step:selection_end} of Algorithm~\ref{alg:datadrivenSMPC}, we obtain the optimal PCE coefficients $\pce{u}^{j,\star}_{0|k}$ of the first input. Notice that $\pcecoe{u}{j,\star}_{0|k} =0 $ holds for $j \in \I_{[2, L-1]}$ due to the causality condition \eqref{eq:causality}. Then, we consider the feedback input $u^{\text{cl}}_k$ as a realization of random variable given by $\pce{u}^{j,\star}_{0|k}$, 
\begin{equation}\label{eq:feedback}
	\textstyle u^{\text{cl}}_k = \sum_{j=0}^{1}\pce{u}^{j,\star}_{0|k} \psi^j(\boldsymbol{\xi}_k\relx) = \pce{u}^{0,\star}_{0|k} +\pce{u}^{1,\star}_{0|k} \xi_k(\omega)
\end{equation}
where  $\xi_k(\omega) \in \R$ is determined by \eqref{eq:projection_Tk}. Observe that, even with the choice of the backup initial condition, \eqref{eq:feedback} is an implicit feedback law computing $u_k^{\text{cl}} $ from $z_k$ that reflects the information of the currently measured data.

\begin{algorithm}[t!]
	\caption{Output-feedback stochastic data-driven predictive control algorithm with OCP \eqref{eq:PCEOCP}} \label{alg:datadrivenSMPC}
	\algorithmicrequire $T$, $N$, $\hat{J}_{N,0} \leftarrow +\infty$, $V^{\text{m}}_N \leftarrow +\infty$,  $V^{\text{b}}_N \leftarrow +\infty$, $k \leftarrow 0$\\
	\textbf{Offline data collection:}
	\begin{algorithmic}[1]
		\State Randomly sample $u_k \in \set{U}$ for $k\in \I_{[0,T-1]}$ 
       \label{step:pre_1}
		\State Apply $\trar{u}{T-1}$ to system~\eqref{eq:ARX} under disturbances $\trar{w}{T-1}$ for  given $(u,y)_{[-\Tini,-1]}$, record  $\trar{y}{T-1}$ 
        \State Record $\trar{w}{T-1}$ or estimate via \eqref{eq:least_square_w}
		\State Determine terminal ingredients ($K$, $P$, $\Gamma$, $\gamma$) with $(u,w,y)_{[0,T-1]}$ \label{step:pre_5}
		\State Construct \eqref{eq:PCEOCP} with $(u,w,y)_{[0,T-1]}$ and $P$, $\Gamma$ and $\gamma$
	\end{algorithmic}
	\begin{algorithmic}[1]
		\Statex\hspace*{-\algorithmicindent}\textbf{Online optimization}
		\State  Measure $z_k=(u,y)_{[k-\Tini,k-1]}$ \label{step:MPC_0}
 		\If{$k \geq 1$} update   $T_k$, $\xi_k\relx$, $\eta_{k-1} \relx$  and $\hat J_{N,k}$
		\EndIf
			\State  Solve OCP~\eqref{eq:PCEOCP} with the measured initial condition~\eqref{eq:consistency_prime} 
			\If{\eqref{eq:PCEOCP} is feasible with optimal cost $V_{N}^{\text{m}} \leq \hat{J}_{N,k}$}
			\State   Apply $u^{\text{cl}}_k =  \pce{u}^{0,\star}_{0|k}$ to system \eqref{eq:ARX} \label{step:selectBegin}	
			\Else 
			\State Solve \eqref{eq:PCEOCP}  with the backup initial condition~\eqref{eq:moment_matching}
			\State Update $V_N^{\text{b}}$ and determine $u^{\text{cl}}_k$ by \eqref{eq:feedback}  
			\State Apply $u^{\text{cl}}_k$  to system \eqref{eq:ARX}
			\EndIf \label{step:selection_end}
			\State $V_{N,k} \leftarrow \min\{V_N^{\text{m}}, V_N^{\text{b}}\}$
			\State  $k\leftarrow k+1$, $V_N^{\text{m}}\leftarrow +\infty$, $V_N^{\text{b}}\leftarrow +\infty$ go to Step \ref{step:MPC_0}
	\end{algorithmic}
\end{algorithm}

\subsection{Analysis of Closed-Loop Properties}\label{sec:stability}
For a specific initial condition $z_{\ini}$ and a specific disturbance realization trajectory $\{w_k\}_{k\in \N}$, we have the following closed-loop dynamics of \eqref{eq:ARX_realization} by applying the feedback input $u_{k}^\text{cl}$ determined by Algorithm~\ref{alg:datadrivenSMPC}
\begin{subequations}\label{eq:ARX_cl}
\begin{equation}\label{eq:ARX_realization_cl}
	y_{k} = \Phi z_{k}  + D u_{k}^\text{cl}  + w_{k}, \quad z_{0}= z_{\ini}.
\end{equation}	
 Moreover, Algorithm~\ref{alg:datadrivenSMPC} generates the sequence $V_{N,k}\in \R $, $k\in \N$,  corresponding to the optimal value functions arising in closed loop.
Accordingly, for a probabilistic initial condition $Z_{\ini}$ and  disturbance $W_k$, $k\in \N$, we have the closed-loop dynamics in random variables
\begin{equation}\label{eq:ARX_cl_RM}
	Y_{k} = \Phi Z_{k} + D U_{k}^\text{cl} +W_{k}, \quad Z_{0}= Z_{\ini}
\end{equation}
\end{subequations}
where conceptually the realization of $U_{k}^\text{cl}$ is $u_{k}^\text{cl}$. Similarly, we define the probabilistic optimal value function $\mcl{V}_{N,k}\in \spl $ as  $\mcl{V}_{N,k}\relx= V_{N,k}$.
The next theorem summarizes the closed-loop properties of the proposed scheme.
\begin{theorem}[Closed-loop properties]\label{thm:stability}
	Let Assumptions~\ref{ass:data}--\ref{ass:Lypunov} hold and let the closed-loop dynamics \eqref{eq:ARX_cl} be determined by Algorithm~\ref{alg:datadrivenSMPC} with perfectly measured disturbance realizations. 
	Suppose that at time instant $k=0$,  OCP~\eqref{eq:PCEOCP} is feasible with the measured initial condition \eqref{eq:consistency_prime}. 
	Then, for 
 \begin{subequations}
 \begin{equation} \label{eq:alpha}
	   \alpha \doteq \trace\left(\Sigma_W(Q+\widetilde{E}^\top P \widetilde{E})\right) \in \R^+
	\end{equation} 
 the following statements hold:
\begin{itemize}
	\item[i)]  For all $k\in\N$, OCP~\eqref{eq:PCEOCP},  with the selection of initial conditions from Algorithm~\ref{alg:datadrivenSMPC}, is feasible.
	\item[ii)] The probabilistic optimal value functions at consecutive time instants satisfy  
\begin{equation}\label{eq:cost_decay}
\mean\left[\mcl{V}_{N,k+1} -\mcl{V}_{N,k}\right] \leq  - \mean \big[ \|U^\text{cl}_{k}\|^2_{R}+ \|Y_{k}\|^2_{Q} \big]+\alpha.
\end{equation}
\item[iii)] The average asymptotic performance is bounded from above by
\begin{equation}\label{eq:average_cost_bound}
\textstyle \lim_{k\rightarrow\infty} \frac{1}{k}\sum_{i=0}^k\mean \big[ \|U^\text{cl}_{i}\|^2_{R}+ \|Y_{i}\|^2_{Q} \big] \leq \alpha.
\end{equation}
\end{itemize}
\end{subequations}
\End
\end{theorem}
The detailed proof is given in Appendix~D. 

We note that recursive feasibility of the OCP (statement i)) does not guarantee the recursive satisfaction of chance constraints in closed loop, which can be interpreted as the recursive satisfaction of one-step conditional probability constraints, i.e., $\mbb P[a^\top_y Y_{k} \leq 1 \,|\, k] \geq 1-\varepsilon$. 
While the one-step chance constraints hold for time instants where measured initial conditions are used, for cases with backup initial conditions, only  $\mbb P[a^\top_y Y_{k} \leq 1 \,|\, {k-\tau}] \geq 1-\varepsilon$ can be ensured, where $k-\tau$, $\tau \in \I_{[0,k-1]}$, is the most recent past time step when the measured initial condition was selected. We refer the readers to \cite[Rem. 1]{Farina2016} for a comprehensive discussion.

Moreover, similar to \cite{Hewing2020,Cannon09m}, statements ii)-iii) utilize the average asymptotic performance bound as a notion of stability for the stochastic closed-loop dynamics. We note that $\alpha$ stands for the asymptotic average cost calculated under $U_k = KZ_k$ with $K$ corresponding to $P$ via \eqref{eq:PK}. Thus, the proposed scheme guarantees either improved or equal asymptotic average cost compared to the feedback policy $U_k = KZ_k$~\cite{Hewing2020}. The equivalence holds when $K$ is determined by a linear-quadratic Riccati design. 

\begin{remark}[Extension to non-i.i.d. disturbances]\label{rem:non_iid}
	If the disturbances have non-zero mean,  one can consider an offset-free data-driven scheme to ensure recursive feasibility and performance guarantees~\cite{Lazar2022}. Conceptually, this offset-free scheme considers the stage cost  $\|Y_k\|_Q^2+\|U_k-U_{k-1}\|_R^2$ to compensate the mean part of the disturbance.  
	For  cases of zero-mean disturbance, if $W_{k+t}$ are independently but not identically distributed for different $t\in \I_{[0,N-1]}$, one can still use the basis \eqref{eq:common_basis}. However, one will obtain different PCE coefficients for each $W_{k+t}$.
	To ensure the closed-loop properties of Theorem~\ref{thm:stability}, one can consider $\Sigma_W$  to be element-wise larger than all $\Sigma[W_{k}]$, $k \in \N$ in this case.
	The consideration of dependent disturbances in the closed-loop analysis remains, however, an open problem. \End
\end{remark}

\subsection{ Data-Driven Design of Terminal Ingredients} \label{sec:terminal}
It remains to discuss the data-driven design of terminal ingredients. 
We first give the data-driven surrogate of $A_K = \widetilde{A} + \widetilde{B} K$ as an extension of the feedback design of \cite{DePersis19}.

Consider a realization trajectory $(u,y)_{[0,T-1]}$ of \eqref{eq:ARX_realization} with given $(u,y)_{[-\Tini,-1]}$ and $w_k$, $k\in \I_{[0,T]}$. For the sake of compact notation, we introduce the following short-hands
\begin{equation}\label{eq:data}
	\begin{aligned}
		\zdd \doteq \begin{bmatrix}
			\hankel_{\Tini}(u_{[-\Tini,T-2]}) \\
			\hankel_{\Tini}(y_{[-\Tini,T-2]})
		\end{bmatrix},&\quad  	\udd \doteq \hankel_1(u_{[0,T-1]}),\\
		\ydd \doteq  \hankel_1(y_{[0,T-1]}),&\quad \wdd \doteq \hankel_1(w_{[0,T-1]})
	\end{aligned}
\end{equation}
with the subscript `$\text{d}$' indicating that they are recorded data. 

\begin{assumption}[Data requirement]\label{ass:sufficient_data} The block matrix
	$
	[\zdd^\top, \udd^\top]^\top	$  $\in \R^{(\dimz+\dimu)\times T}
	$
	has full row rank. \End
\end{assumption}
 For LTI systems without disturbance, a sufficient condition for Assumption~\ref{ass:sufficient_data} to hold is $\Tini\cdot \dimy = \dimx$, where $\dimx$ is the minimal state dimension, see \cite[Lemma 13]{Berberich2021t}. 
This condition requires the system to admit a specific structure. Hence,  Assumption~\ref{ass:sufficient_data} is relatively restrictive when no disturbance is considered. However, in the presence of disturbances, $(\widetilde A,[\widetilde B, \widetilde E])$ in \eqref{eq:extended_dyna} is a controllable pair by Lemma~\ref{lem:controllable}. This implies that a sufficient condition for  Assumption~\ref{ass:sufficient_data} is that $(\udd,\wdd)$ is persistently exciting of order $\dimz + 1$ as shown in \cite[Cor. 2]{Willems2005}. 
  Therefore,  the full rank condition of Assumption~\ref{ass:sufficient_data} is less restrictive than $\Tini\cdot \dimy = \dimx$. 
%Section~\ref{sec:examples} draws upon a numerical example to illustrate this.

\begin{lemma}[Data-driven surrogate of $A_K$]\label{lem:data}
	Let Assumption~\ref{ass:sufficient_data} hold. Then, for any given $K\in \R^{\dimu \times \dimz}$, there exists a $H \in \R^{T \times\dimz}$ such that
		\begin{equation}\label{eq:IK1}
			[	\zdd^\top,\udd^\top]^\top
			H = [ 
			I_{\dimz},
			K^\top
			]^\top.
		\end{equation} 
		The data-driven reformulation of $A_K=\widetilde{A} + \widetilde{B} K$ reads
		\begin{equation*}
			A_K= SH , S \doteq \begin{bmatrix}
				\bar{A}	\zdd +\bar{B}\udd \\
				\ydd-\wdd
			\end{bmatrix}  \in \R^{\dimz \times T}. \tag*{\End}
		\end{equation*}
\end{lemma}
\vspace*{3pt}
\begin{proof}
	Observe that \eqref{eq:IK1} holds given the full row rank condition of Assumption~\ref{ass:sufficient_data}. Horizontally stacking the data of the realization dynamics in \eqref{eq:ARX_realization}, we have 
	$
	\ydd =\left[\Phi~D\right] [\begin{smallmatrix}
		\zdd \\ 
		\udd
	\end{smallmatrix}] +\wdd.
	$
	We arrive at
	$
	A_K=[\widetilde{A}~\widetilde{B}][\begin{smallmatrix}
		I_{\dimz}\\ 
		K
	\end{smallmatrix}] =[\begin{smallmatrix}
		\bar A & \bar B\\
		\Phi & D 
	\end{smallmatrix}][\begin{smallmatrix}
		\zdd\\ 
		\udd
	\end{smallmatrix}]H= SH.
	$
\end{proof}

One key difference between the above result and the approach of \cite{DePersis19} is that the above utilizes the known matrix components $\bar{A}$ and $\bar{B}$ appearing in \eqref{eq:extended_dyna}, while  \cite{DePersis19} treats the system matrices $[\widetilde{A}~\widetilde{B}]$ as unknown.

With the data-driven surrogate $SH$ of $A_K$, we turn to the data-driven construction of $K$, $P$, and $\Gamma$.
Observe that the Lyapunov equality \eqref{eq:PK} is equivalent to 
\begin{align*}
	A_K^\top (P + \widetilde{Q})A_K - (P + \widetilde{Q}) &=  -\left(K^\top R K + \widetilde{Q} \right)
\end{align*}
with $\widetilde{Q}\doteq \widetilde{E}Q \widetilde{E}^\top$. Since this resembles the closed-loop Lyapunov equation of an LQR design, we choose $K$ as the LQR feedback of the state-space model~\eqref{eq:extended_dyna}. \vspace*{2pt}

For stabilizable $(\widetilde{A}, \widetilde{B})$ the LQR design for \eqref{eq:extended_dyna} is equivalent to solving \cite{Doerfler2021}
\begin{subequations}\label{eq:solveK}
	\begin{align}
		\min_{
			\substack{
				\widetilde{P} \in \R^{\dimz\times\dimz},K\in \R^{  \dimu\times\dimz}
			}
		} &\trace(\widetilde{Q}\widetilde{P}+K^\top R K  \widetilde{P}) \\
		\text{s.t. }\quad
		\widetilde{P}& \succeq \widetilde{E}\widetilde{E}^\top\\
		(\widetilde{A}+\widetilde{B}K)&\widetilde{P}(\widetilde{A}+\widetilde{B}K)^\top - \tilde{P} +\widetilde{E}\widetilde{E}^\top\preceq 0.
	\end{align}
\end{subequations}
Replacing $\widetilde{A}+\widetilde{B}K= SH$ and following the convexification procedure outlined  in \cite{Doerfler2021}, a data-driven and tractable reformulation of \eqref{eq:solveK} reads
\begin{subequations}\label{eq:terminal_tractable}
	\begin{align}
		\min_{
			\substack{
				X_1\in \R^{\dimu\times\dimu},X_2\in \R^{T \times \dimz}
			}
		}\trace(\widetilde{Q} \zdd X_2)&+\trace(X_1) \\
		\text{s.t. }\quad
		\begin{bmatrix}
			\zdd X_2 -\widetilde{E}\widetilde{E}^\top &S X_2   \\
			\star  & 	\zdd X_2
		\end{bmatrix}&\succeq 0\\
		\begin{bmatrix}
			X_1 &  R^{\frac{1}{2}} \udd X_2 \\
			\star & 
			\zdd X_2
		\end{bmatrix}&\succeq  0
	\end{align}
\end{subequations}
with $K = \udd X_2 (\zdd X_2)^{-1}$ and $H = X_2 (\zdd X_2)^{-1} $.

Using $K,H$  we obtain $P$ and  $\Gamma$ by solving
\begin{subequations}\label{eq:PK_Gamma_data}
	\begin{gather}
		H^\top S^\top( P + \widetilde{Q}) S H- P  =  -K^\top R K,  \\
		H^\top S^\top \Gamma S H- \Gamma  =  -I_{\dimz}.
	\end{gather}
	We compute $\gamma$ using \eqref{eq:gamma}. Finally, we specify the terminal constraint on the expected value $\Zf$
	as a sublevel set
	  of the terminal cost, i.e.,
	$
	\Zf \doteq\{z\in \R^{\dimz}\,|\,  \|z\|_P^2 \leq \varepsilon_z\}
	$ with a sufficiently small $\varepsilon_z\in\R^+$ such that  Assumption~\ref{ass:Lypunov} holds.
\end{subequations}

\subsection{Robustness Analysis with Estimated Disturbances}\label{sec:estimate_disturbance}
	The previous discussions are based on the assumption of exact measurement of past disturbances. Next, we first recap a disturbance estimation approach from \cite{Pan21s}, and then we analyze the closed-loop behavior of Algorithm~\ref{alg:datadrivenSMPC} with estimated past disturbances.

	Due to \cite[Prop. 2]{Pan21s} the measured data $\udd$, $\ydd$, $\zdd$, and $\wdd$ in \eqref{eq:data} satisfies  $ 
	(\ydd - \wdd)( I_T-[\begin{smallmatrix}
		\zdd\\ \udd
	\end{smallmatrix}]^\rinv [\begin{smallmatrix}
		\zdd\\ \udd
	\end{smallmatrix}]) = 0$.
    Henceforth,   we estimate the past disturbance sequence via the least-square solution to the previous equation, i.e., we use 
\begin{equation}\label{eq:least_square_w}
\hat{\boldsymbol{w}}_\text{d} \doteq	\hankel_1(\hat{ w}_{[0,T-1]})  = \ydd \left( I_T-\begin{bmatrix}
		\zdd\\ \udd
	\end{bmatrix}^\rinv\begin{bmatrix}
		\zdd\\ \udd
	\end{bmatrix}\right).
\end{equation}

\begin{lemma}[Data-driven one-step representation]\label{lem:dataARX} 
    Let Assumption~\ref{ass:sufficient_data} hold. Then, system~\eqref{eq:ARX_realization} has the following equivalent representation
    \begin{subequations}
    \begin{equation}\label{eq:ddModel}
    y_{k} =	(\boldsymbol{y}_{\text{d}} - \boldsymbol{w}_{\text{d}}) \begin{bmatrix}
    	\zdd\\ \udd
    \end{bmatrix}^\rinv \begin{bmatrix}
    	z_k\\
    	u_k
    \end{bmatrix}+w_{k}.
    \end{equation}
    Moreover, replacing $\boldsymbol{w}_{\text{d}}$ with its least-square estimate $\hat{\boldsymbol{w}}_\text{d}$ from \eqref{eq:least_square_w}, we have the one step prediction
    \begin{equation}\label{eq:estimated_model}
    y^{\text p}_{k} = \boldsymbol{y}_{\text{d}}\begin{bmatrix}
    	\zdd\\ \udd
    \end{bmatrix}^\rinv \begin{bmatrix}
    	z_k\\
    	u_k
    \end{bmatrix}+w_{k} .
    \end{equation}
    \end{subequations} \End
\end{lemma}
\begin{proof}
    The proof of \eqref{eq:ddModel} follows arguments similar to the state feedback case in \cite[Thm. 1]{DePersis19}. Moreover,  with $\boldsymbol{w}_{\text{d}}$ replaced by its least-square estimate $\hat{\boldsymbol{w}}_\text{d}$, we have
$       y^{\text p}_{k} =( \boldsymbol{y}_{\text{d}}-  \hat{\boldsymbol{ w }}_{\text{d}})[\begin{smallmatrix}
			\zdd\\ \udd
		\end{smallmatrix}]^\rinv [\begin{smallmatrix}
			z_k\\
			u_k
		\end{smallmatrix}]+w_{k} \stackrel{\eqref{eq:least_square_w}}{=}\boldsymbol{y}_{\text{d}}[\begin{smallmatrix}
			\zdd\\ \udd
		\end{smallmatrix}]^\rinv [\begin{smallmatrix}
			z_k\\
			u_k
		\end{smallmatrix}]+w_{k}.$
\end{proof}

Notice that the Hankel matrix equations \eqref{eq:RVfunda}--\eqref{eq:mixed_funda} of Lemma~\ref{lem:RVfundamental} hold  for $(u,y,\hat{w})_{[0,T-1]}$ combined with the estimated $\hat{ w}_{[0,T-1]}$. They characterize the system dynamics of \eqref{eq:estimated_model}, cf. \cite[Cor. 3]{Pan21s}. We define the underlying model mismatch corresponding to the disturbance estimate $\hat{\boldsymbol{w}}_\text{d}$ as 
\[
\Delta\doteq \wdd \begin{bmatrix}
			\zdd\\ \udd
		\end{bmatrix}^\rinv \in \R^{\dimy \times (\dimz + \dimu)}.
\]

\begin{assumption}[Bounded optimal solution] \label{ass:bound}
	For all $\pce{\bar z}^{[0,L-1]}_k$  for which OCP~\eqref{eq:PCEOCP} is feasible, the set of  optimal solutions % $(\pce{  u},\pce{ y})^{[0,L-1],\star}_{[0,N-1]|k}$ 
	$(\seq{\pce{u}}_k^\star,\seq{\pce{y}}_k^\star)$
	of OCP~\eqref{eq:PCEOCP} is non-empty and bounded. \End
\end{assumption}
We note that the above assumption can be satisfied by constraining the decision variables within a compact set.

 \begin{proposition}[Robustness of performance]\label{cor:stability}
 Let the conditions of Theorem \ref{thm:stability} and Assumption~\ref{ass:bound} hold.
  Consider OCP~\eqref{eq:PCEOCP} with $ w_{[0,T-1]}$ replaced by $\hat{w}_{[0,T-1]}$ from \eqref{eq:least_square_w}. 
 At time instant $k=0$, suppose that OCP~\eqref{eq:PCEOCP} is feasible with the measured initial condition \eqref{eq:consistency_prime}. Then,  statement i) of Theorem~\ref{thm:stability} holds and there exist $C \in \R^+$ such that  
\begin{equation}\label{eq:cost_decay_2}
  \begin{aligned}
    \mean[ \mcl{V}_{N,k+1} -\mcl{V}_{N,k}]  \leq & - \mean \left[ \|U^\text{cl}_{k}\|^2_{R}+ \|Y_{k} - \Delta  [\begin{smallmatrix}
  	Z_k\\
  	U_k
  \end{smallmatrix}]\|^2_{Q} \right] \\
   &+\alpha + \mean \left[C\left\|\Delta[\begin{smallmatrix}
   	Z_k\\
   	U_k
   \end{smallmatrix}]\right\| \right].
\end{aligned}  
\end{equation}
		\End 
	\end{proposition}
The detailed proof is given in Appendix~E. 
We remark that \eqref{eq:cost_decay_2} recovers \eqref{eq:cost_decay} when there is no model mismatch, i.e., when  $\Delta \to 0$.

\section{Numerical Example}\label{sec:examples}
We consider an LTI aircraft model exactly discretized with sampling time $t_s=0.5~\text{s}$ from \cite{maciejowski02predictive}. The ARX matrices are	
		\begin{align*}
			\Phi &= \scalemath{0.7}{\begin{bmatrix} -0.019	&-\phantom{0}1.440	&-\phantom{0}0.201&	\phantom{-}0.256&	0.050	&\phantom{0}0.160&	-\phantom{0}0.256&	0.086\\
				\phantom{-}0.711&	-\phantom{0}1.800&	-\phantom{0}4.773&	\phantom{0}3.688&	0.650&	\phantom{0}2.982&	-\phantom{0}2.688&	1.707\\
				\phantom{-}1.444&	-26.922	&-15.746&	12.898&	2.319	&10.461	&-12.897	&5.171
			\end{bmatrix}},
		\end{align*}
and	$D = 0_{3\times 1}$ with $\dimy=3$, $\dimu=1$, $T_\ini=2$, and thus $\dimz=8$. A minimal state-space representation with $\dimx=4$ is given in \cite{Pan21s}.
We consider the ARX dynamics with data as above, where $W_k, k\in \N$, are i.i.d. random variables from Gaussian mixture models, as the simulated plant. The underlying Gaussian mixture model is a mixture of $\mcl{N}\big([-0.005,-0.5,-0.05], \diag([0.0001,1,0.01])\big)$ and $\mcl{N}\big([0.005,0.5,0.05], \diag([0.0001,1,0.01])\big)$ with zero mean. Despite its name, Gaussian mixture models are not Gaussian distributions. They admit infinite support, which renders recursive feasibility and robust constraint satisfaction challenging.
The corresponding
$\pce M_w = \left[\begin{smallmatrix} 0.011 & 0.002 & 0.002 \\
0.002 & 1.118 & 0.020 \\
0.002 & 0.020 & 0.110
\end{smallmatrix}\right]$ in \eqref{eq:ARX_PCE} is determined as the principle square root of the covariance of the disturbance.
Note that $Y^j$ denotes the $j$-th element of $Y$.
We impose  chance constraints $\mathbb{P}[Y^1_{i|k} \geq -1] \geq 1- \varepsilon_y
$ and $\mathbb{P}[Y^1_{i|k} \leq 1] \geq 1- \varepsilon_y$ for all $i \in \I_{[0,N-1]}$. We choose $ \varepsilon_y = 0.1$ which corresponds to $\sigma(\varepsilon_y)=3$.
The weighting matrices in the stage cost are $Q=\diag([1,1,1])$ and $R = 1$. 

We compare two schemes: I) Algorithm \ref{alg:datadrivenSMPC} with disturbance measurement. II) Algorithm \ref{alg:datadrivenSMPC} with disturbance estimation.
In the data collection phase of Scheme I), we record an input-output-disturbance trajectory of 1000 steps; while in Scheme II), only the input-output trajectory is recorded and further used to estimate the disturbance realizations via \eqref{eq:least_square_w}. 

We use the first 22 recorded input-output data pairs and the measured/estimated disturbances to determine the terminal ingredients $P$, $\Gamma$, $\gamma$, and $\Zf$ for the two schemes by the data-driven procedure outlined in Section~\ref{sec:terminal}. 
Observe that $\Tini\cdot \dimy = \dimx$ does not hold for this system since $2 \cdot 3\neq 4$ while the generated data satisfies Assumption~\ref{ass:sufficient_data}. For the sake of comparison, we also generate a 22-step input-output trajectory with the same inputs and initial condition without disturbance. We observed that for this data  Assumption~\ref{ass:sufficient_data} is not satisfied. This indicates that in the presence of disturbances, the data requirement condition of Assumption~\ref{ass:sufficient_data} is less restrictive than in the disturbance-free case. 

To solve \eqref{eq:terminal_tractable} for $K$ and $H$ we use \texttt{MOSEK}~\cite{mosek}. Compared to the feedback $K^\star$ obtained by directly solving the usual discrete-time algebraic Riccati equation for the exact system matrices of \eqref{eq:extended_dyna}, the relative difference is $\|K-K^\star\|_2^2\cdot(\|K^\star\|_2^2)^{-1}$ $=0.088$ for Scheme II), while it is $ 1.01\cdot 10^{-3}$ for Scheme I). 
Note that for both cases $A_K$ is Schur stable. Finally, we calculate $P$ and $\Gamma$ solving~\eqref{eq:PK_Gamma_data} and thus obtain $\gamma$ via \eqref{eq:gamma}.

We apply Algorithm~\ref{alg:datadrivenSMPC} with prediction horizon $N=10$ using \texttt{IPOPT} \cite{waechter06}. The computations are done on a virtual machine with an AMD EPYC processer with 2.8 GHz, 32GB of RAM in \texttt{julia}. Similar to before, in the data collection phase we use the first 90 recorded inputs-outputs and measured/estimated disturbances to construct the Hankel matrices. Applying Lemma~\ref{lem:finite_PCE} we obtain the PCE expression for each component of $W_k$. Moreover, the dimension of the overall PCE basis is $L= 1+ n_\xi + N n_\eta = 32$  and the number of decision variables for the considered OCP with null-space projection is 3488 \cite{Pan21s}.

We sample 1000 different disturbance realization sequences of length 31 each and then compute the corresponding closed-loop responses of the two schemes using distributed computing in \texttt{julia}. The computation times, the average asymptotic costs, and the highest empirical relative frequency of constraint violations $\max_k \mathbb{P}^\text{l}_k$, $\max_k \mathbb{P}^\text{u}_k$ are summarized in Table \ref{tab:AircraftComparisonTime}.
 Here, the  empirical relative frequency of constraint violation $\mathbb{P}^\text{l}_k$ represents the percentage of time instants where $y^1_{k} < -1$ over all $1000$ realizations at time instant $k$, and $\mathbb{P}^\text{u}_k$ is defined similarly for $y_k^1>1$. 
Due to the distributionally robust formulation of \eqref{eq:PCEOCP}, the highest empirical relative frequencies of constraint violation for Schemes~I) and II) is substantially lower than the allowed violation probability of 10$\%$.
The two schemes exhibit similar results when compared based on the mean and the Standard Deviation (SD) of the computation time of each time step evaluated in the closed loop.

 Moreover, evaluating the closed-loop realization trajectories over time, we obtain the average cost $\bar{\ell}=\mean[ \sum_{k=11}^{30}\ell(k)]/20 $ with $\ell(k) =\|u_k\|_R^2+\|y_k\|_Q^2$. The two schemes achieve similar average costs.  Moreover, since $K$ is determined from data-driven LQR design as in Section \ref{sec:terminal}, the theoretical value of the average asymptotic cost for Scheme~I) is $\alpha = 540.94$ by Theorem \ref{thm:stability}, which is consistent with the simulation results of Scheme I) with $\bar{\ell} = 537.99$ and Scheme II) with $\bar{\ell} = 543.54$.

\begin{figure}[t!]
	\begin{center}
		\includegraphics[width=1\linewidth,trim={60mm 20mm 40mm 30mm},clip]{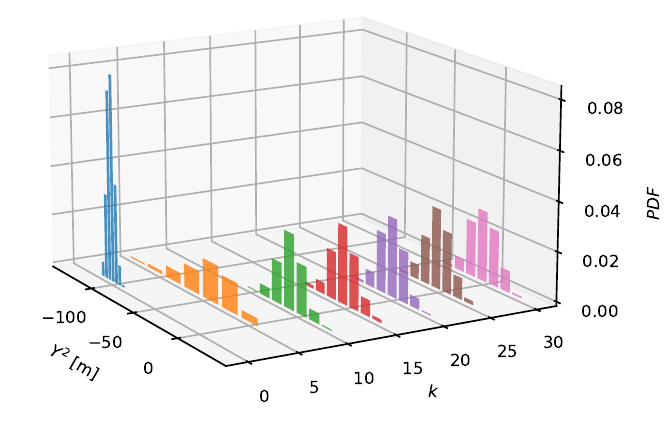}
		\caption{Histograms of the output $Y^2$ from 1000 closed-loop realization trajectories of Scheme II).} 
		\label{fig:AircraftDistEvolution}
	\end{center}
\end{figure}

\begin{figure}[t!]
	\begin{center}
		\includegraphics[width=8.4cm]{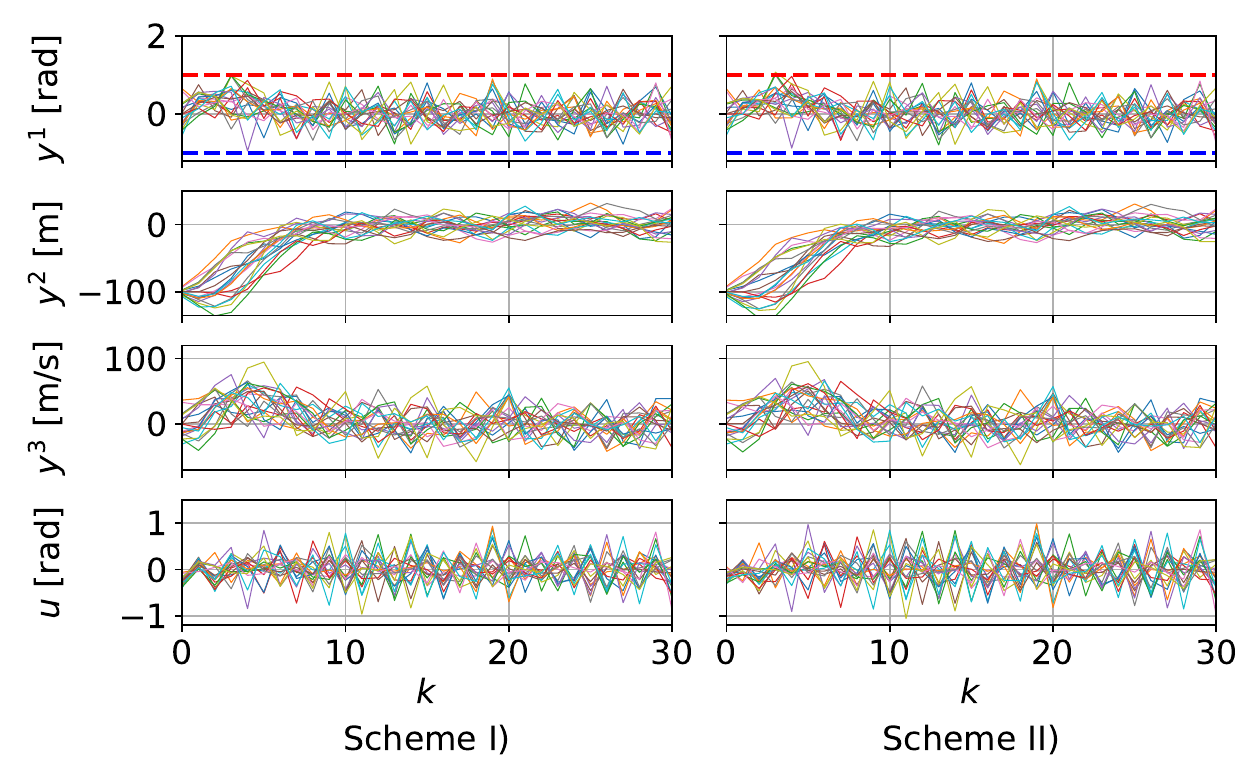}
		\caption{20 different closed-loop realization trajectories of Algorithm~\ref{alg:datadrivenSMPC}. Blue-dashed line: $y^1=-1 $; red-dashed line: $y^1=1$. Left: Scheme I) with measured disturbances; right: Scheme II) with estimated disturbances. \label{fig:AircraftTraj}} 		
	\end{center}
\end{figure}

Additionally, for Scheme II), the time evolution of the (normalized) histograms of the output realizations $y^2$ at $k=0, 5,10, 15, 20, 25, 30$ is depicted in Figure~\ref{fig:AircraftDistEvolution}, where the vertical axis refers to the (approximated) probability density of $Y^2$. As one can see, the proposed scheme controls the system to a stationary distribution of~$Y^2$ centered at $0$.
With $20$ different sampled sequences of disturbance realizations, we show the corresponding closed-loop realization trajectories of Schemes I) and II) in Figure~\ref{fig:AircraftTraj}. It can be seen that the both schemes exhibit similar closed-loop responses.

\begin{table}[t!]
	\caption{Comparison of the computation times, the average costs, and the number of violations for 1000 realized closed-loop trajectories with a length of 31 steps  for measured and estimated disturbances for Schemes I) and II).}
	\label{tab:AircraftComparisonTime}
	\centering
     \begin{adjustbox}{width=\columnwidth}
		\begin{tabular}{cccccc}
			\toprule
			\multirow{2}{*}{\shortstack{Scheme} }&  \multicolumn{2}{c}{Computation time} 
			& 	\multirow{2}{*}{$\bar{\ell} $} &\multirow{2}{*}{$\displaystyle \max_k \mathbb{P}^\text{l}_k$} & \multirow{2}{*}{$\displaystyle \max_k \mathbb{P}^\text{u}_k$}\\ 
			\cmidrule(lr){2-3} 
			 & Mean $\SI{}{[s]}$ & SD $\SI{}{[s]}$ & & & \\
			\midrule
			I & 0.66 &  0.050  & 537.99 & 0.3\%, $k =11$  & 3.1\%, $k =3$\\
			II & 0.68 & 0.049 & 543.54 & 0.2\%, $k =\phantom{0}7$ & 4.1\%, $k =3$\\
			\bottomrule
		\end{tabular}
	\end{adjustbox}
\end{table}

\section{Conclusions}\label{sec:conclusion}
This paper has presented and analyzed a data-driven output-feedback predictive control scheme for stochastic LTI systems. Based on a stochastic variant of Willems' fundamental lemma---which relies on polynomial chaos expansions---we presented sufficient conditions for recursive feasibility and a characterization of the closed-loop performance. Moreover, we have discussed the data-driven design of terminal ingredients which satisfy the proposed conditions.

It deserves to be noted that we only require knowledge of the first two moments but not exact knowledge of the underlying distribution. Thus, the proposed scheme considers a weak form of distributional robustness and it is applicable to a wide class of Gaussian and non-Gaussian settings. A numerical example has demonstrated the efficacy of the proposed scheme for Gaussian mixture models. 
Future work will consider improving the robustness, extending to nonlinear systems, and using the regularization techniques for measurement noise from \cite{Coulson2019,Berberich20}.  

~\\
\section*{Appendix}

\subsection{Proof of Lemma~\ref{lem:controllable}}
	\begin{proof}
		We examine the controllability of the pair $(\widetilde{A},\left[\widetilde{B},\widetilde{E}\right])$ via the rank of the reachability matrix  \[\mathcal{R}(\widetilde{A},[\widetilde{B},\widetilde{E}]) = \left[ \widetilde A^{\lag-1}[\widetilde{B},\widetilde{E}],\widetilde A^{\lag-2}[\widetilde{B},\widetilde{E}],\cdots , [\widetilde{B},\widetilde{E}]\right].\] 
		
		Observe that for $i\in \I_{[0,\lag-1]}$,
		\[
		\widetilde A^i  \widetilde{B}  = \begin{bmatrix}
			0_{(\lag-1-i)\dimu \times \dimu} \\
			I_{n_u} \\
			0_{ n_u i\times n_u} \\
			\star_{\lag n_y\times n_u} 
		\end{bmatrix},\quad \widetilde A^i  \widetilde{E}  = \begin{bmatrix}
			0_{\lag n_u\times n_y}  \\
			0_{(\lag-1-i)\dimy \times \dimy} \\
			I_{n_y} \\
			\star_{ n_y i\times n_y} \\
		\end{bmatrix},
		\]
		where $\star$ indicates matrix blocks  depending on $\Phi$ and $D$.
		 Observe that when multiplied by $\widetilde A^i$, the identity matrix blocks $I_{\dimu}$ in $\widetilde{B}$ and $I_{\dimy}$ in $\widetilde{E}$ are shifted upwards by $\dimu i$ rows and $\dimy i$ rows, respectively. As a result,
		\begin{align*}
			\mathcal{R}(\widetilde A,\widetilde B)  &= \left[\begin{matrix}
				I_{\lag \dimu} \\ 
				\star
			\end{matrix} \right],
			\mathcal{R}(\widetilde A, \widetilde E) &= \left[\begin{smallmatrix}
				0_{\lag \dimu\times\dimy} & 0 &0 \\
				I_{\dimy}  &0 &0 \\
				\vdots  &   \ddots & 0\\
				\star & \star  & I_{\dimy}\\
			\end{smallmatrix}\right].
		\end{align*}
		Therefore, $\mcl{R}(\widetilde A,[\widetilde B,\widetilde E])$, which possesses the same row rank as $\left[\mcl{R}(\widetilde A,\widetilde B),\mcl{R}(\widetilde A, \widetilde E)\right]$, is of full row rank. Hence $(\widetilde A,[\widetilde B, \widetilde E])$ is a  controllable pair.
	\end{proof}

\subsection{Proof of Lemma~\ref{lem:finite_PCE}}
    \begin{proof}
	 Since $ \pce M \in \R^{\dimv \times n_\xi}$ is in general not invertible, we utilize  the Moore-Penrose inverse to construct $\xi$. We note that the Moore-Penrose inverse $ \pce M^\dagger$ of $ \pce M \in \R^{\dimv \times n_\xi}$ is unique and satisfies the following properties, cf. \cite{Penrose1955}, 
	 \begin{subequations}
	 \begin{align}
\pce	M \pce M^\dagger \pce M = \pce M &, \pce M^\dagger \pce M  \pce M^\dagger=\pce M^\dagger \label{eq:MP_inv}\\
	\pce M  \pce M^\dagger =  (\pce M\pce  M^\dagger)^\top&, \pce M^\dagger \pce  M=  (\pce M^\dagger \pce M)^\top. \label{eq:MP_sym}
\end{align}
	 \end{subequations}
	Let 
	\begin{equation}\label{eq:xi_construction}
		\xi= \pce M^\dagger (V-\mean[V]) + (I_{n_\xi} - \pce M^\dagger \pce M) \xi_{\mcl N} 
	\end{equation}
	with $\xi_{\mcl N} \in \mcl{D}(0,I_{n_\xi})$ independent of $V$. We note that one could choose any random variable from $\splx{n_\xi}$ with unit covariance and zero mean.   Specifically, if $\pce M$ is of full column rank, then \eqref{eq:xi_construction} recovers $\xi= \pce M^\dagger (V-\mean[V])$ since  in this case $ \pce M^\dagger \pce M= I_{n_\xi}$.

	Next, we prove that  $\xi$ from \eqref{eq:xi_construction} satisfies the required conditions:  $\mean[\xi] = 0$,  $\Sigma[\xi] = I_{n_\xi}$, and $V = \mean[V] + \pce M \xi $. First, we have
$
		\mean[\xi] =\pce M^\dagger (\mean[V]-\mean[V]) + (I_{n_\xi} - \pce M^\dagger \pce M)\mean[\xi_{\mcl N}] = 0$ since $\mean[\xi_{\mcl N}] = 0$ by construction. Furthermore,
	\begin{align*}
		\Sigma[\xi]& = \pce M^\dagger \Sigma[V]\pce M^{\dagger \top} + (I_{n_\xi} - \pce M^\dagger \pce M) \Sigma [\xi_{\mcl N} ](I_{n_\xi} - \pce M^\dagger \pce M)^\top\\
		&=\pce M^\dagger (\pce M \pce M^\top) \pce M^{\dagger \top}+  (I_{n_\xi} - \pce M^\dagger  \pce M) (I_{n_\xi} - \pce M^\dagger \pce M)^\top\\
		&		= \pce M^\dagger \pce M (\pce M^\dagger \pce M)^\top + I_{n_\xi} - \pce M^\dagger \pce  M\\
	   & \quad -  (\pce M^\dagger \pce M)^\top + \pce M^\dagger \pce M  (\pce M^\dagger \pce M)^\top  \stackrel{\eqref{eq:MP_sym}}{=} I_{n_\xi}.
	\end{align*}
We note that the random-variable equality $V = \mean[V] + \pce M \xi $ is considered as $V\relx =  \mean[V] + \pce M \xi \relx$ for $\mu$-almost all $\omega \in \Omega$. 
Let $ \Delta \doteq V - \mean[V] - \pce M \xi $. This equality is  equivalent to $\Delta\relx = 0$  for $\mu$-almost all $\omega \in \Omega$, which is true iff $\mean[\Delta] = 0 $ and $\mean[\Delta^\top \Delta] = 0$ \cite[Problem 3 (v)]{fristedt13modern}. 
With $\xi$ from \eqref{eq:xi_construction}, we have
	\begin{align*}
	 &	\Delta= V - \mean[V] - \pce M \big(\pce M^\dagger (V-\mean[V]) + (I_{n_\xi} - \pce M^\dagger \pce M) \xi_{\mcl N} \big) \\
		& =  (I_{\dimv} - \pce M \pce M^\dagger)(V - \mean[V])- (\pce M - \pce M \pce M^\dagger \pce M)\xi_{\mcl N}\\
		& \stackrel{\eqref{eq:MP_inv}}{=} (I_{\dimv} - M M^\dagger)(V - \mean[V]).
	\end{align*}
Hence, we have $\mean[\Delta] = (I_{\dimv} - \pce M \pce M^\dagger)\mean[(V - \mean[V])] =0$ and
	\begin{align*}
	&	\mean[\Delta^\top \Delta ] 
		= \trace\left((I_{\dimv} - \pce M \pce M^\dagger)^\top  (I_{\dimv} - \pce M \pce M^\dagger) \Sigma[V] \right)\\
		& =  \trace\left( (I_{\dimv} -\pce M \pce M^\dagger)^\top  (I_{\dimv} -\pce M \pce M^\dagger) ( \pce M \pce M^\top)  \right) \\
		&= \trace\left((I_{\dimv} - \pce M \pce M^\dagger)^\top \left(\pce M \pce M^\top -\pce M \pce M^\dagger \pce M \pce M^\top  \right)\right)  = 0,
	\end{align*}
where the last equality holds since $  \pce M (\pce M^\dagger \pce M \pce M^\top)  \stackrel{\eqref{eq:MP_inv}}{=} \pce M \pce M^\dagger$.

Combining the above results, we conclude that there exists $\xi$ given by \eqref{eq:xi_construction} that satisfies the conditions of Lemma~\ref{lem:finite_PCE}.
\end{proof}
	
\subsection{Proof of Proposition~\ref{pro:recur}}
This proof is organized into two main steps.  As a preparatory step, we give a technical lemma that shows the recursive feasibility of OCP~\eqref{eq:PCEOCP} using an extended PCE basis.  Then, we turn to the proof of Proposition~\ref{pro:recur} by using projection $T_k$ to keep the dimension of the basis constant.

\begin{lemma}[Recursive feasibility with extended basis]\label{lemma:recurFeasi}
	Let Assumptions~\ref{ass:data}--\ref{ass:Lypunov} hold. At time instant $k-1$,  given the basis $\{\psi^j(\boldsymbol{\xi}_{k-1})\}_{j=0}^{L-1}$ as in \eqref{eq:common_basis},
 suppose that OCP~\eqref{eq:PCEOCP}
 is feasible with the initial condition   $\bar{ \pce{z}}^{[0,L-1]}_{k-1}$  where $\bar{ \pce{z}}^{j}_{k-1} =0$, $\forall j \in \I_{[1+n_\xi,L-1]}$.

At time instant $k$,  considering the  extended basis
\begin{equation}\label{eq:basis_app}
	\begin{aligned}
		\{\psi^j(\tilde{\boldsymbol{\xi}}_k)\}_{j=0}^{\widetilde{L}-1} 	,\,
		\tilde{\boldsymbol{\xi}}_k =  [\xi_{k-1}^\top,\eta_{[k-1,k+N-1]}^\top]^\top
	\end{aligned}
\end{equation} with $ \widetilde{L}=L+n_\eta = 1+n_\xi+(N+1)n_\eta$ and $\psi^j$ from \eqref{eq:P1}, OCP~\eqref{eq:PCEOCP} is feasible with the  initial condition
		\begin{align}
				\bar{\pce{z}}^{[0,n_\xi + n_\eta]}_{k} = %\begin{cases} 
					\pce{z}^{[0,n_\xi + n_\eta],\star}_{1|k-1}, \quad					\bar{\pce{z}}^{[n_\xi + n_\eta+1,\widetilde{L}-1]}_{k} = 	0
				 \label{eq:backupIni} 
			\end{align} 
		where $\pce{z}^{j,\star}_{1|k-1}$ is the predicted optimal solution of $\pce{z}^{j}_k$ at time instant $k-1$.
\End 
\end{lemma}
\begin{proof} 
	Consider the optimal solution at $k-1$, i.e. $(\pce{u},\pce{y},\pce{z})^{j,\star}_{t|k-1}$ for $j\in \I_{[0,L-1]}$ and $t \in \I_{[0, N-1]}$.
	 Similar to the proof of recursive feasibility in deterministic MPC, we construct a candidate feasible solution of OCP~\eqref{eq:PCEOCP} at time instant $k$ by shifting the last optimal solution based on the original basis \eqref{eq:common_basis}.
For all $j \in \I_{[0,L-1]}$, we have
\begin{subequations}\label{eq:feasibleShift}
	\begin{align}
		\{\tilde{\pce{u}}^j_{t|k}\}_{t=0}^{N-2} & = \{\pcecoe{u}{j,\star}_{t|k-1}\}_{t=1}^{N-1}, & \tilde{\pce{u}}^j_{N-1|k} &= K \pcecoe{z}{j,\star}_{N|k-1}\\
		\{\tilde{\pce{y}}^j_{t|k}\}_{t=0}^{N-2} & = \{\pcecoe{y}{j,\star}_{t|k-1}\}_{t=1}^{N-1}, & \tilde{\pce{y}}^j_{N-1|k} &=\widetilde{E}^\top \tilde{\pce{z}}^j_{N|k}
		\\
		\{\tilde{\pce{z}}^j_{t|k} \}_{t=0}^{N-1} &= \{ \pcecoe{z}{j,\star}_{t|k-1}\}_{t=1}^{N}, & \tilde{\pce{z}}^j_{N|k} &= A_K \pcecoe{z}{j,\star}_{N|k-1} \label{eq:shiftedX} 
	\end{align}
and	for all  $j \in \I_{[L,\widetilde{L}-1]}$ we have 
	\begin{align}
	 \{\tilde{\pce{u}}^j_{t|k}\}_{t=0}^{N-1} = 0,\, \{\tilde{\pce{y}}^j_{t|k}\}_{t=0}^{N-2} = 0,\, \{\tilde{\pce{z}}^j_{t|k}\}_{t=0}^{N-1} = 0, \label{eq:shiftedAfterL}\\
		\tilde{\pce{y}}^j_{N-1|k} =\pce{w}^j_{k+N-1},\quad \tilde{\pce{z}}^j_{N|k} =\widetilde{E}\pce{w}^j_{k+N-1}. 
	\end{align}
\end{subequations}
Note that as the last disturbance involves new independent stochastic uncertainty, the above shifted trajectories of PCE coefficients are expressed in the extended basis \eqref{eq:basis_app} with $\tilde{L} = L+n_\eta$ with the corresponding input coefficients being set to $0$ in \eqref{eq:shiftedAfterL}.

Next, we prove the feasibility of \eqref{eq:feasibleShift} in OCP~\eqref{eq:PCEOCP} with respect to the extended basis \eqref{eq:basis_app}. Thanks to the shift construction of \eqref{eq:feasibleShift} and Assumption~\ref{ass:Lypunov}, constraints \eqref{eq:H_PCE_SOCP_hankel}--\eqref{eq:PCEOCP_chancey} are satisfied. For the terminal constraints \eqref{eq:terminal}, we see $\tilde{\pce{z}}^0_{N|k} = A_K \pcecoe{z}{0,\star}_{N|k-1}$ $\in \Zf$ holds for $\pcecoe{z}{0,\star}_{N|k-1}\in \Zf$, cf. Assumption~\ref{ass:Lypunov}.  For the terminal constraint on the covariance, we have
\begin{equation*}
	\begin{aligned}
		\textstyle{\sum_{j=1}^{\widetilde{L}-1}}&\|\tilde{\pce{z}}^{j}_{N|k} \|_\Gamma^2 = 	\textstyle{\sum_{j=1}^{L-1}} \pce{z}^{j,\star\top}\pred{N}{k-1} \underbrace{	A_K^\top \Gamma A_K}_{ \stackrel{\eqref{eq:Gamma}}{=}\,\Gamma-I_{\dimz} } \pce{z}^{j,\star}\pred{N}{k-1}  \\
		&+ \textstyle{\sum_{j=L}^{\widetilde{L}-1}} \pce{w}^{j\top}_{k+N-1}\widetilde{E}^\top \Gamma \widetilde{E}\pce{w}^{j}_{k+N-1}
		\\
		  \leq& \left( \textstyle 1-\frac{1}{\lambda_{\text{max}}(\Gamma)}\right)\textstyle{\sum_{j=1}^{L-1} }\|\pce{z}^{j,\star}\pred{N}{k-1}\|_{\Gamma}^2   
		 +\trace(\Sigma_W\widetilde{E}^\top \Gamma \widetilde{E}) \\
	 \leq 	& \left(\textstyle 1-\frac{1}{\lambda_{\text{max}}(\Gamma)}\right) \gamma +\trace(\Sigma_W\widetilde{E}^\top \Gamma \widetilde{E})	 \stackrel{\eqref{eq:gamma}}{=} \gamma. \\
	\end{aligned} 
\end{equation*} 
Thus, we conclude that at time instant $k$, OCP~\eqref{eq:PCEOCP} constructed with the extended basis~$\{\psi^j(\tilde{\boldsymbol{\xi}}_k)\}_{j=0}^{\widetilde{L}-1}$ is feasible when \eqref{eq:backupIni} is used as initial condition, and \eqref{eq:feasibleShift} is a feasible solution.
\end{proof}

Now we give the proof of Proposition~\ref{pro:recur}~\\
\begin{proof}
 We construct a feasible solution candidate using the projection matrix $T_k$.
    According to Lemma~\ref{lem:RVfundamental}, for each $j\in \I_{[0,\widetilde{L}-1]}$, there exists  $\tilde{\pce{g}}^{j}$ corresponding to  $\{(\tilde{\pce{u}},\tilde{\pce{y}})^{j}_{t|k}\}_{t=-T_\ini}^{N-1}$  as shown in \eqref{eq:feasibleShift} such that the Hankel equality constraint~\eqref{eq:H_PCE_SOCP_hankel} holds. Since the initial conditions are related by the projection $T_k$ in \eqref{eq:moment_matching}, we consider $\hat{\pce{g}}^0 = \tilde{\pce{g}}^0$,
    	\begin{align*}
    	\hankel_1\left(\hat{\pce{g}}^{[1,n_\xi]}\right) &= \hankel_1\left(\tilde{\pce{g}}^{[1,n_\xi+n_\eta]}\right) T_k, 	\\ \hat{\pce{g}}^{[n_\xi+1,L-1]} &= \tilde{\pce{g}}^{[n_\xi+n_\eta+1,\tilde{L}-1]}.
    	\end{align*}
    Next, we exploit  Lemma~\ref{lem:RVfundamental} again. That is, for each $j \in \I_{[0,L-1]}$,  the resulting $\{(\hat{\pce{u}},\hat{\pce{y}})^{j}_{t|k}\}_{t=-T_\ini}^{N-1}$ of evaluating \eqref{eq:H_PCE_SOCP_hankel} with $\hat{\pce{g}}^{j}$ is an input-output trajectory of the PCE coefficient dynamics. For all $\pce{v} \in \{\pce{u},\pce{y},\pce{z}\}$ and  all $t \in\I_{[0,N-1]}$, they satisfy  $\hat{\pce{v}}_{t|k}^j = \tilde{\pce{v}}_{t|k}^j$, 
    	\begin{equation}\label{eq:projection}
    		\begin{aligned}
    	    \hankel_1\left(\hat{\pce{v}}^{[1,n_\xi]}_{t|k}\right) &= \hankel_1\left(\tilde{\pce{v}}^{[1,n_\xi+n_\eta]}_{t|k}\right)	T_k, \\	\hat{\pce{v}}_{t|k}^{[n_\xi+1,L-1]}& = \tilde{\pce{v}}_{t|k}^{[n_\xi+n_\eta+1,\tilde{L}-1]}.
    		\end{aligned}
    	\end{equation}
    In other words, the resulting $\{(\hat{\pce{u}},\hat{\pce{y}})^{j}_{t|k}\}_{t=-T_\ini}^{N-1}$  satisfy \eqref{eq:H_PCE_SOCP_hankel}--\eqref{eq:PCEOCP_u}.
    Turning to the chance constraints \eqref{eq:PCEOCP_chanceu}--\eqref{eq:PCEOCP_chancey}, we note that for all $i \in \I_{[1,N_u]}$, $t\in\I_{[0, N-1]}$, we have
    \begin{align*}
    & \|a^\top_{u,i} \hankel_1\left(\hat{\pce{u}}^{[1,L-1]}_{t|k}\right)\|^2 -\|a^\top_{u,i}\hankel_1\left(\tilde{\pce{u}}^{[1,L+n_\eta-1]}_{t|k}\right)\|^2\\
    & = -  \|a^\top_{u,i} \hankel_1\left(\hat{\pce{u}}^{[1,n_\xi+n_\eta]}_{t|k}\right)\|^2_{ (I_{n_\xi+n_\eta}-T_k T_k^\top)} \leq 0,
    \end{align*}
    as  $I_{n_\xi+n_\eta}-T_k T_k^\top$ is positive semi-definite since $T_k \in \R^{(n_\xi+n_\eta)\times n_\xi}$ is orthonormal.
    This holds similarly for the PCE coefficients of outputs. Therefore, applying the projection of the feasible solution \eqref{eq:feasibleShift} by \eqref{eq:projection} the left-hand sides of  \eqref{eq:PCEOCP_chanceu}--\eqref{eq:PCEOCP_chancey} do not increase. Thus, the chance constraints are satisfied with $\{(\hat{\pce{u}},\hat{\pce{y}})^{j}_{t|k}\}_{t=-T_\ini}^{N-1}$.
    Similarly, applying \eqref{eq:projection} the left-hand side of the terminal constraint  \eqref{eq:terminal} also does not increase. Thus, we conclude the feasibility of OCP with respect to the basis \eqref{eq:common_basis} and the initial condition specified per \eqref{eq:moment_matching}.
\end{proof}

\subsection{Proof of Theorem~\ref{thm:stability}}

\begin{proof}
    Note that assertion i) on recursive feasibility directly follows from Proposition~\ref{pro:recur} and the procedure of initial condition selection. 
    In the following, we first prove the closed-loop value function decay condition \eqref{eq:cost_decay} (assertion ii)) and then use it to establish the average asymptotic performance bound \eqref{eq:average_cost_bound} (assertion iii)).

 \noindent \textbf{Assertion ii):}   To show that the closed-loop value function decay between time instant $k$ and $k+1$ results from the selection of the initial condition, we rely on the optimal solution at time instant $k$, i.e. $(\pce{u},\pce{y})^{[0,L-1],\star}_{[0,N-1]|k}$ and its shifted trajectories  $(\tilde{\pce{u}},\tilde{\pce{y}})^{[0,\tilde L-1]}_{[0,N-1]|k+1}$  from \eqref{eq:feasibleShift}.  Furthermore, using \eqref{eq:projection_Tk} and \eqref{eq:projection}, we obtain the candidate solution $(\pce{ \hat u},\pce{\hat y})^{[0,L-1]}_{[0,N-1]|k+1}$ for the OCP with the backup initial condition~\eqref{eq:moment_matching}. 
 For the sake of readability, we omit the subscripts  $\cdot|k+1$ and  $\cdot|k$.
 
Consider the resulting performance  of the candidate solution  
  \begin{align*}
  \hat {J}_{N,k+1}  \doteq \textstyle \sum_{t =0}^{N-1} (\hat \ell_{y,t} + \hat \ell_{u,t})
  + \hat \ell_{z,N} 
 \end{align*}
by evaluating \eqref{eq:obj_reformulation}.
We show that    $\hat {J}_{N,k+1}$ can be expressed using the shifted solution $(\tilde{\pce{u}},\tilde{\pce{y}})$. We first link the part $\hat \ell_{y,t}$  to  $\tilde{\pce{y}}$, with $n_\xi =1$,
\begin{align*}
& \hat \ell_{y, t}  \doteq  \textstyle \|\pcecoe{\hat y}{0}_{i} + \pcecoe{\hat y}{1}_{i} \xi_{k+1}\relx \|_Q^2 +  \sum_{j= 2 }^{L-1}   \|\pcecoe{\hat y}{j}_{i} \|_Q^2\\
& \textstyle \stackrel{\eqref{eq:projection}}{=}   \|\pcecoe{\tilde y}{0}_{i} + \hankel_1(\pcecoe{\tilde y}{[1,1+n_\eta]}_{i}) T_k T_k^\top [\begin{smallmatrix}
		\xi_k \relx \\
		\eta_k \relx
	\end{smallmatrix}]  \|_Q^2 +   \sum_{j= 2+n_\eta }^{L+n_\eta-1}   \|\pcecoe{\tilde y}{j}_{i} \|_Q^2
\\
& \textstyle \stackrel{\eqref{eq:projection_Tk}}{=}  \|\pcecoe{\tilde y}{0}_{i} + \hankel_1(\pcecoe{\tilde y}{[1,1+n_\eta]}_{i}) [\begin{smallmatrix}
		\xi_k \relx \\
		\eta_k \relx
	\end{smallmatrix}]\|_Q^2 +  \sum_{j= 2+n_\eta }^{L+n_\eta-1}   \|\pcecoe{\tilde y}{j}_{i} \|_Q^2.
\end{align*}
At time instant $k$, we do not know $\eta_k \relx$ but  its mean $\mean[\eta_k] =0$ and covariance $\Sigma[\eta_k] = I_{n_\eta}$. Denote $\mean[\hat{\ell}_{y, t} | k ]$ as the conditional expectation of $\hat{\ell}_{y, t} $ at time instant $k$  given $\xi_k\relx$ but with unknown $\eta_k \relx$.
Let  
$\tilde y_t = \pcecoe{\tilde y}{0}_{t} + \pcecoe{\tilde y}{1}_{t} \xi_k \relx$, we have
\begin{align}
	&\mean[\hat{\ell}_{y,t} | k ] = \mean\left[  \|\tilde y_t + \hankel_1( \pcecoe{\tilde y}{[2,1+n_\eta]}_t) \eta_k  \|_Q^2\right]  +  \textstyle\sum_{j= 2+n_\eta }^{L+n_\eta-1}   \|\pcecoe{\tilde y}{j}_{i} \|_Q^2 \nonumber \\
	& = \|\tilde y_t \|_Q^2 + \trace\left( \hankel_1(\pcecoe{\tilde y}{[2,1+n_\eta]}_t)^\top Q   \hankel_1(\pcecoe{\tilde y}{[2,1+n_\eta]}_t) \Sigma[\eta_k]  \right)  \nonumber \\
	& \textstyle + 2 \tilde y^\top_t Q   \hankel_1(\pcecoe{\tilde y}{[2,1+n_\eta]}_t) \mean [\eta_k]  + \sum_{j= 2+n_\eta }^{L+n_\eta-1}   \|\pcecoe{\tilde y}{j}_{t} \|_Q^2 \nonumber \\
	&  = \|\pcecoe{\tilde y}{0}_{t} + \pcecoe{\tilde y}{1}_{t} \xi_k \relx\|_Q^2 + \textstyle \sum_{j=2}^{L+n_\eta-1}   \|\pcecoe{\tilde y}{j}_{t} \|_Q^2 \doteq \tilde \ell_{y,t}. \label{eq:mean_ellk}
\end{align}
Similarly, the above also holds for the other parts of $\hat {J}_{N,k+1}$, i.e., $ \mean[\hat \ell_{u,t}\,|\,k ]= \tilde \ell_{u,t}  $ and $ \mean[\hat \ell_{z,N}\,|\,k ]= \tilde \ell_{z,N} $.
As a result, we have
\[
\mean[\hat {J}_{N,k+1}\,|\,k]  = \textstyle \sum_{t=0}^{N-1} (\tilde{\ell}_{y,t} +\tilde{\ell}_{u,t}) +\tilde{\ell}_{z,N}.
\]
Moreover, we have $V_{N,k+1}\leq \hat J_{N,k+1}$ from \eqref{eq:costdecay}. Thus, we have $\mean [\mcl V_{N,k+1}\,|\, k]\leq \mean[\hat {J}_{N,k+1}\,|\,k]$.

Now, we have the upper bound for    $ \mean [\mcl V_{N,k+1}\,|\, k] -  V_{N,k}$ 
\begin{align*}
	&  \mean [\mcl V_{N,k+1}\,|\, k] -  V_{N,k}	 \leq \mean[\hat {J}_{N,k+1}\,|\,k] - V_{N,k}  \\
	& =\tilde \ell_{y,{N-1}}+ \tilde \ell_{u,{N-1}}+\tilde \ell_{z,{N}}  -\ell^{\star}_{u,0} - \ell^{\star}_{y,0}   -  \ell_{z,{N}}^\star
\end{align*} 
where the identical terms $\{\tilde \ell_{u,t},\tilde \ell_{y,t}\}_{t=0}^{N-2}$ and  $\{\ell_{u,t}^\star,\ell_{y,{t}}^\star\}_{t=1}^{N-1}$ are cancelled.
Furthermore,
\begin{align*}
&\tilde \ell_{y,{N-1}} + \tilde  \ell_{u,{N-1}} + \tilde  \ell_{z,{N}} -  \ell_{z,{N}}^\star\\
 &=  \textstyle  \|\pcecoe{\tilde y}{0}_{N-1} + \pcecoe{\tilde y}{1}_{N-1} \xi_k \relx\|_Q^2 +  \sum_{j=2}^{L-1}   \|\pcecoe{\tilde y}{j}_{N-1} \|_Q^2 \tag{$i$} \\
&+   \textstyle \|\pcecoe{\tilde u}{0}_{N-1} + \pcecoe{\tilde u}{1}_{N-1} \xi_k \relx\|_R^2 +  \sum_{j=2}^{L-1}   \|\pcecoe{\tilde u}{j}_{N-1} \|_R^2 \tag{$ii$} \\
&+    \textstyle \|\pcecoe{\tilde z}{0}_{N} + \pcecoe{\tilde z}{1}_{N} \xi_k \relx\|_P^2 +  \sum_{j=2}^{L-1}   \|\pcecoe{\tilde z}{j}_{N} \|_P^2  \tag{$iii$}\\
&-   \textstyle \|\pcecoe{ z}{0,\star}_{N} + \pcecoe{ z}{1,\star}_{N} \xi_k \relx\|_P^2 -  \sum_{j=2}^{L-1}   \|\pcecoe{z}{j,\star}_{N-1} \|_P^2 \tag{$iv$}\\
& \,\textstyle + \sum_{j=L}^{L+n_\eta -1} \left(\|\pcecoe{\tilde y}{j}_{N-1}\|_Q^2 + \| \pcecoe{\tilde u}{j}_{N-1}\|_R^2+ \|\pcecoe{\tilde z}{j}_N\|_P^2\right), \tag{$v$}
\end{align*}
Exploiting \eqref{eq:feasibleShift}, we rewrite the terms $(i)+(ii)+(iii)$ and $(v)$  
	\begin{gather*}
	(i)+(ii)+(iii)= \textstyle \|\pcecoe{ z}{0,\star}_{N} + \pcecoe{ z}{1,\star}_{N} \xi_k \relx\|_\Xi^2 +  \sum_{j=2}^{L-1}   \|\pcecoe{z}{j,\star}_{N-1} \|_\Xi^2    ,\\
		\Xi= K^\top R K +A_K^\top\widetilde{E}^\top Q\widetilde{E}A_K +A_K^\top PA_K \stackrel{\eqref{eq:PK}}{=}P,\\
		(v)\stackrel{\eqref{eq:feasibleShift}}{=}	\sum_{j=L}^{\tilde{L}-1}\|\pce{w}^{j}_{k+N}\|_{Q +  \widetilde{E}^\top P\widetilde{E}}^2 = \trace(\Sigma_W (Q +  \widetilde{E}^\top P \widetilde{E})) = \alpha. 
	\end{gather*}
Hence,  $(i)+(ii)+(iii) + (iv) = 0$ and $(v) = \alpha$. 

Combining the previous results, we arrive at
	\begin{equation*}
	     \begin{aligned}
		&\mean [\mcl V_{N,k+1}\,|\, k] -  V_{N,k}	  \leq -\ell^{\star}_{u,0} - \ell^{\star}_{y,0} +\alpha\\  
		&=- \left( \|\pce{u}^{0,\star}_{0} + \pce{u}^{1,\star}_{0} \xi_k \relx  \|_R^2+\|\pce{y}^{0,\star}_{0} + \pce{y}^{1,\star}_{0} \xi_k \relx \|_Q^2\right)  +\alpha \\
	&	=- \|u_{k}^{\text{cl}}\|_R^2 -  \mean\left[ \|Y_{k}\|^2_Q \,\middle | \,k \right] + \alpha.
	\end{aligned}
	\end{equation*}
As the above inequality holds for each closed-loop implementation, we lift  $V_{N,k+1} -  V_{N,k}$  to its random variable counterpart and obtain \eqref{eq:cost_decay}.
 
 \noindent \textbf{Assertion iii):} 
	Recursively using \eqref{eq:cost_decay} from $0$ to $k$ and let $k\to \infty$, we have
	\begin{align*}
	0\leq&  \lim_{k\rightarrow\infty} \frac{1}{k} \mean\left[\mcl{V}_{N,k} -\mcl{V}_{N,0}\right] =\lim_{k\rightarrow\infty} \frac{1}{k} {\sum_{i=0}^k} (\mean\left[\mcl{V}_{N,i+1} -\mcl{V}_{N,i}\right])\\
	\leq & \displaystyle \alpha- \frac{1}{k}\lim_{k\rightarrow\infty}\sum_{i=0}^k	  \mean \left[ \|U^\text{cl}_{i}\|^2_{R}+ \|Y_{i}\|^2_{Q} \right].
	\end{align*}
	Thus, we obtain the average performance condition \eqref{eq:average_cost_bound} for the closed-loop system in random variables \eqref{eq:ARX_cl_RM}.
\end{proof}
 
\subsection{Proof of Proposition~\ref{cor:stability}}
\begin{proof}
	At time instant $k$, given $z_k$, $u_k^{cl}$, and $w_k$, the model mismatch between \eqref{eq:ddModel} and \eqref{eq:estimated_model} leads to a prediction error of $y_k$ as $ \delta_k = y_k - y^\text{p}_k = \Delta [z_k^\top,u^{\text{cl}\top}_k ]^\top.$
By appending $y_k$ and $y^\text{p}_k$ to previously recorded inputs and outputs, we obtain
\begin{align*}
	z_{k+1}& = [u_{[k-\Tini+2,k+1]}^{\text{cl}\top},y_{[k-\Tini+2,k]}^\top, y^{\top}_{k}]^\top, \\
	z^\text{p}_{k+1} & = [u_{[k-\Tini+2,k+1]}^{\text{cl}\top},y_{[k-\Tini+2,k]}^\top, y^{\text{p}\top}_{k}]^\top,
\end{align*}
which leads to different 	$\eta_k \relx$ and 	$\eta_k^\text{p} \relx$ by \eqref{eq:eta_estimation}. Note that 	$\mean[\eta_k^\text{p}] = 0$ and $\Sigma[\eta_k^\text{p}] = I_{n_\eta}$, and
	\begin{align*}
		\eta_k -\eta_k^\text{p} & = \pce M_w^\dagger \widetilde{E}^\top (z_{k+1} - z_{k+1}^\text{p}) = \pce M_w^\dagger \delta_k, \\
		\mean[\eta_k] &= \mean[\eta_k^\text{p}] + \eta_k -\eta_k^\text{p} = \eta_k -\eta_k^\text{p}= \pce M_w^\dagger \delta_k, \\
	\Sigma[\eta_k] &= \Sigma[\eta_k^\text{p} + \eta_k -\eta_k^\text{p}] = \Sigma[\eta_k^\text{p}]  = I_\eta.
	\end{align*}
Due to $\mean[\eta_k] \neq \mean[\eta_k^\text{p}] = 0$, \eqref{eq:mean_ellk} in the proof of Theorem~\ref{thm:stability} does not hold but
\begin{align*}
\mean[\hat{\ell}_{y,t} | k ] = \tilde{\ell}_{y,t}  + 2 \tilde y^\top_t Q   \hankel_1(\pcecoe{\tilde y}{[2,1+n_\eta]}_t) \pce M_w^\dagger \delta_k .
\end{align*}
Note that the shifted solution  $(\pce{ \tilde u},\pce{\tilde y})$ is constructed from the optimal solution of the last time instant. Given the optimal solution  lives in a bounded set due to Assumption~\ref{ass:bound}, there exists $ C_{ y,t} \in\R^+$ such that 
\[
\mean[\hat{\ell}_{y,t} | k ] \leq \tilde{\ell}_{y,t} + C_{ y,t}\|\delta_k\|.
\]
Subsequently, there exist $C_{ y,t}, C_{ u,t} \in\R^+$ for $i \in \I_{[0,N-1]}$, $C_{ z,N} \in \R^+$, and $C \in \R^+$ such that
\begin{align*}
&\textstyle \mean[\hat J _{N,k+1}\,|\, k] \leq \sum_{i=0}^{N-1} (\tilde{\ell}_{y,i} +\tilde{\ell}_{u, i}) +\tilde{\ell}_{z, N} \\
&\textstyle +   (\sum_{i=0}^{N-1} (C_{ y,t} + C_{ u,t}) + C_{z,N}) \|\delta_k\| \doteq \tilde J_{N,k+1}  + C \|\delta_k\|.
\end{align*}
Finally, we arrive at
    \begin{align*}
    	&\mean [V_{N,k+1}\,|\, k] -  V_{N,k} \leq \mean[\hat{J}_{N,k+1} \,| \,k] - V_{N,k}\\
     & \leq    \tilde J_{N,k+1} - V_{N,k} + C \|\delta_k\| \\
     &	= - \|u_{k}^{\text{cl}}\|_R^2 -  \mean\left[ \|Y_{k}^{\text{p}}\|^2_Q \,\middle | \,k \right] +\alpha+ C \|\delta_k\| \\
     & =  - \|u_{k}^{\text{cl}}\|_R^2 -  \mean\left[ \|Y_{k} - \delta_k\|^2_Q \,\middle | \,k \right] +\alpha +C \|\delta_k\|.
    \end{align*}
	    	As the above inequality holds for all  $\omega \in \Omega$, we lift it to its random variable counterpart and thus arrive at \eqref{eq:cost_decay_2}.
\end{proof}

\section*{Acknowledgment}
The authors acknowledge the very constructive and helpful suggestions made by the anonymous reviewers and by the associate editor.

\bibliographystyle{IEEEtran}
\bibliography{IEEEabrv}

\begin{thebibliography}{10}
\providecommand{\url}[1]{#1}
\csname url@samestyle\endcsname
\providecommand{\newblock}{\relax}
\providecommand{\bibinfo}[2]{#2}
\providecommand{\BIBentrySTDinterwordspacing}{\spaceskip=0pt\relax}
\providecommand{\BIBentryALTinterwordstretchfactor}{4}
\providecommand{\BIBentryALTinterwordspacing}{\spaceskip=\fontdimen2\font plus
\BIBentryALTinterwordstretchfactor\fontdimen3\font minus
  \fontdimen4\font\relax}
\providecommand{\BIBforeignlanguage}[2]{{%
\expandafter\ifx\csname l@#1\endcsname\relax
\typeout{** WARNING: IEEEtran.bst: No hyphenation pattern has been}%
\typeout{** loaded for the language `#1'. Using the pattern for}%
\typeout{** the default language instead.}%
\else
\language=\csname l@#1\endcsname
\fi
#2}}
\providecommand{\BIBdecl}{\relax}
\BIBdecl

\bibitem{Willems2005}
J.~C. Willems, P.~Rapisarda, I.~Markovsky, and B.~L.~M. De~Moor, ``A note on
  persistency of excitation,'' \emph{Systems \& Control Letters}, vol.~54,
  no.~4, pp. 325--329, 2005.

\bibitem{DePersis19}
C.~De~Persis and P.~Tesi, ``Formulas for data-driven control: Stabilization,
  optimality, and robustness,'' \emph{IEEE Transactions on Automatic Control},
  vol.~65, no.~3, pp. 909--924, 2019.

\bibitem{Markovsky21r}
I.~Markovsky and F.~D{\"o}rfler, ``Behavioral systems theory in data-driven
  analysis, signal processing, and control,'' \emph{Annual Reviews in Control},
  vol.~52, pp. 42--64, 2021.

\bibitem{Yang15}
H.~Yang and S.~Li, ``A data-driven predictive controller design based on
  reduced {Hankel} matrix,'' in \emph{2015 10th Asian Control Conference
  (ASCC)}.\hskip 1em plus 0.5em minus 0.4em\relax IEEE, 2015, pp. 1--7.

\bibitem{Coulson2019}
J.~Coulson, J.~Lygeros, and F.~D{\"o}rfler, ``Data-enabled predictive control:
  In the shallows of the {DeePC},'' in \emph{2019 18th European Control
  Conference (ECC)}.\hskip 1em plus 0.5em minus 0.4em\relax IEEE, 2019, pp.
  307--312.

\bibitem{Berberich20}
J.~Berberich, J.~K{\"o}hler, M.~A. M{\"u}ller, and F.~Allg{\"o}wer,
  ``Data-driven model predictive control with stability and robustness
  guarantees,'' \emph{IEEE Transactions on Automatic Control}, vol.~66, no.~4,
  pp. 1702--1717, 2020.

\bibitem{Doerfler2023}
F.~Dörfler, J.~Coulson, and I.~Markovsky, ``Bridging direct and indirect
  data-driven control formulations via regularizations and relaxations,''
  \emph{IEEE Transactions on Automatic Control}, vol.~68, no.~2, pp. 883--897,
  2023.

\bibitem{Carlet2022}
P.~G. Carlet, A.~Favato, R.~Torchio, F.~Toso, S.~Bolognani, and F.~Dörfler,
  ``Real-time feasibility of data-driven predictive control for synchronous
  motor drives,'' \emph{IEEE Transactions on Power Electronics}, vol.~38,
  no.~2, pp. 1672--1682, 2023.

\bibitem{Bilgic22}
D.~Bilgic, A.~Koch, G.~Pan, and T.~Faulwasser, ``Toward data-driven predictive
  control of multi-energy distribution systems,'' \emph{Electric Power Systems
  Research}, vol. 212, p. 108311, 2022.

\bibitem{Wang2022a}
J.~Wang, Y.~Zheng, K.~Li, and Q.~Xu, ``{DeeP-LCC: Data-enabled predictive
  leading cruise control in mixed traffic flow},'' \emph{IEEE Transactions on
  Control Systems Technology}, vol.~31, no.~6, pp. 2760--2776, 2023.

\bibitem{Berberich2021t}
J.~Berberich, J.~K{\"o}hler, M.~A. M{\"u}ller, and F.~Allg{\"o}wer, ``{On the
  design of terminal ingredients for data-driven MPC},''
  \emph{IFAC-PapersOnLine}, vol.~54, no.~6, pp. 257--263, 2021.

\bibitem{Bongard2022}
J.~Bongard, J.~Berberich, J.~Köhler, and F.~Allgöwer, ``Robust stability
  analysis of a simple data-driven model predictive control approach,''
  \emph{IEEE Transactions on Automatic Control}, vol.~68, no.~5, pp.
  2625--2637, 2023.

\bibitem{Kerz21d}
S.~Kerz, J.~Teutsch, T.~Brüdigam, M.~Leibold, and D.~Wollherr, ``Data-driven
  tube-based stochastic predictive control,'' \emph{IEEE Open Journal of
  Control Systems}, vol.~2, pp. 185--199, 2023.

\bibitem{Wang2022}
Y.~Wang, K.~You, D.~Huang, and C.~Shang, ``Data-driven output prediction and
  control of stochastic systems: An innovation-based approach,''
  \emph{Automatica}, vol. 171, p. 111897, 2025.

\bibitem{Farina2015}
M.~Farina, L.~Giulioni, L.~Magni, and R.~Scattolini, ``An approach to
  output-feedback {MPC} of stochastic linear discrete-time systems,''
  \emph{Automatica}, vol.~55, pp. 140--149, 2015.

\bibitem{Pan21s}
G.~Pan, R.~Ou, and T.~Faulwasser, ``On a stochastic fundamental lemma and its
  use for data-driven optimal control,'' \emph{IEEE Transactions on Automatic
  Control}, vol.~68, no.~10, pp. 5922--5937, 2023.

\bibitem{Faulwasser2022}
T.~Faulwasser, R.~Ou, G.~Pan, P.~Schmitz, and K.~Worthmann, ``Behavioral theory
  for stochastic systems? {A} data-driven journey from {Willems to Wiener} and
  back again,'' \emph{Annual Reviews in Control}, vol.~55, pp. 92--117, 2023.

\bibitem{Pan2023a}
G.~Pan, R.~Ou, and T.~Faulwasser, ``Towards data-driven stochastic predictive
  control,'' \emph{International Journal of Robust and Nonlinear Control}, pp.
  1--23, 2023.

\bibitem{Pan2023}
G.~Pan and T.~Faulwasser, ``Distributionally robust uncertainty quantification
  via data-driven stochastic optimal control,'' \emph{IEEE Control Systems
  Letters}, vol.~7, pp. 3036--3041, 2023.

\bibitem{fristedt13modern}
B.~E. Fristedt and L.~F. Gray, \emph{A Modern Approach to Probability
  Theory}.\hskip 1em plus 0.5em minus 0.4em\relax Springer Science \& Business
  Media, 2013.

\bibitem{Oldewurtel2008}
F.~Oldewurtel, C.~N. Jones, and M.~Morari, ``A tractable approximation of
  chance constrained stochastic {MPC} based on affine disturbance feedback,''
  in \emph{2008 47th IEEE conference on decision and control}.\hskip 1em plus
  0.5em minus 0.4em\relax IEEE, 2008, pp. 4731--4736.

\bibitem{Drgona2020}
J.~Drgoňa, J.~Arroyo, I.~Cupeiro~Figueroa, D.~Blum, K.~Arendt, D.~Kim, E.~P.
  Ollé, J.~Oravec, M.~Wetter, D.~L. Vrabie, and L.~Helsen, ``All you need to
  know about model predictive control for buildings,'' \emph{Annual Reviews in
  Control}, vol.~50, pp. 190--232, 2020.

\bibitem{Sadamoto2022}
T.~Sadamoto, ``On equivalence of data informativity for identification and
  data-driven control of partially observable systems,'' \emph{IEEE
  Transactions on Automatic Control}, vol.~68, no.~7, pp. 4289--4296, 2023.

\bibitem{Hewing2020}
L.~Hewing, K.~P. Wabersich, and M.~N. Zeilinger, ``Recursively feasible
  stochastic model predictive control using indirect feedback,''
  \emph{Automatica}, vol. 119, p. 109095, 2020.

\bibitem{Goulart2006}
P.~J. Goulart, E.~C. Kerrigan, and J.~M. Maciejowski, ``Optimization over state
  feedback policies for robust control with constraints,'' \emph{Automatica},
  vol.~42, no.~4, pp. 523--533, 2006.

\bibitem{farina13probabilistic}
M.~Farina, L.~Giulioni, L.~Magni, and R.~Scattolini, ``A probabilistic approach
  to model predictive control,'' in \emph{2013 52nd IEEE Conference on Decision
  and Control (CDC)}.\hskip 1em plus 0.5em minus 0.4em\relax IEEE, 2013, pp.
  7734--7739.

\bibitem{Witsenhausen68}
H.~S. Witsenhausen, ``A counterexample in stochastic optimum control,''
  \emph{SIAM Journal on Control}, vol.~6, no.~1, pp. 131--147, 1968.

\bibitem{Witsenhausen71}
------, ``Separation of estimation and control for discrete time systems,''
  \emph{Proceedings of the IEEE}, vol.~59, no.~11, pp. 1557--1566, 1971.

\bibitem{Mitter99}
S.~Mitter and A.~Sahai, ``Information and control: Witsenhausen revisited,'' in
  \emph{Learning, control and hybrid systems}.\hskip 1em plus 0.5em minus
  0.4em\relax Springer, 1999, pp. 281--293.

\bibitem{wiener38homogeneous}
N.~Wiener, ``The homogeneous chaos,'' \emph{American Journal of Mathematics},
  pp. 897--936, 1938.

\bibitem{sullivan15introduction}
T.~J. Sullivan, \emph{{Introduction to Uncertainty Quantification}}.\hskip 1em
  plus 0.5em minus 0.4em\relax Springer, 2015, vol.~63.

\bibitem{paulson14fast}
J.~A. Paulson, A.~Mesbah, S.~Streif, R.~Findeisen, and R.~D. Braatz, ``Fast
  stochastic model predictive control of high-dimensional systems,'' in
  \emph{53rd IEEE Conference on Decision and Control}.\hskip 1em plus 0.5em
  minus 0.4em\relax IEEE, 2014, pp. 2802--2809.

\bibitem{Mesbah14}
A.~Mesbah, S.~Streif, R.~Findeisen, and R.~D. Braatz, ``Stochastic nonlinear
  model predictive control with probabilistic constraints,'' in \emph{2014
  American Control Conference (ACC)}.\hskip 1em plus 0.5em minus 0.4em\relax
  IEEE, 2014, pp. 2413--2419.

\bibitem{Ou21}
R.~Ou, M.~H. Baumann, L.~Gr{\"u}ne, and T.~Faulwasser, ``A simulation study on
  turnpikes in stochastic {LQ} optimal control,'' \emph{IFAC-PapersOnLine},
  vol.~54, no.~3, pp. 516--521, 2021, 16th IFAC Symposium on Advanced Control
  of Chemical Processes ADCHEM 2021.

\bibitem{muehlpfordt18comments}
T.~M{\"u}hlpfordt, R.~Findeisen, V.~Hagenmeyer, and T.~Faulwasser, ``Comments
  on quantifying truncation errors for polynomial chaos expansions,''
  \emph{IEEE Control Systems Letters}, vol.~2, no.~1, pp. 169--174, 2018.

\bibitem{Witteveen2006}
J.~A.~S. Witteveen and H.~Bijl, ``Modeling arbitrary uncertainties using
  {Gram-Schmidt} polynomial chaos,'' in \emph{44th AIAA Aerospace Sciences
  Meeting and Exhibit}, N.~J. Pfeiffer, Ed.\hskip 1em plus 0.5em minus
  0.4em\relax United States: American Institute of Aeronautics and Astronautics
  Inc. (AIAA), 2006, p. 896.

\bibitem{GhanSpan03}
R.~G. Ghanem and P.~D. Spanos, \emph{{Stochastic Finite Elements: A Spectral
  Approach}}, revised~ed.\hskip 1em plus 0.5em minus 0.4em\relax Springer New
  York, 2003.

\bibitem{Calafiore2006}
G.~C. Calafiore and L.~E. Ghaoui, ``On distributionally robust
  chance-constrained linear programs,'' \emph{Journal of Optimization Theory
  and Applications}, vol. 130, no.~1, pp. 1--22, 2006.

\bibitem{Lobo1998}
M.~S. Lobo, L.~Vandenberghe, S.~Boyd, and H.~Lebret, ``Applications of
  second-order cone programming,'' \emph{Linear algebra and its applications},
  vol. 284, no. 1-3, pp. 193--228, 1998.

\bibitem{Hewing18s}
L.~Hewing and M.~N. Zeilinger, ``Stochastic model predictive control for linear
  systems using probabilistic reachable sets,'' in \emph{2018 IEEE CDC}.\hskip
  1em plus 0.5em minus 0.4em\relax IEEE, 2018, pp. 5182--5188.

\bibitem{Farina2016}
M.~Farina and R.~Scattolini, ``Model predictive control of linear systems with
  multiplicative unbounded uncertainty and chance constraints,''
  \emph{Automatica}, vol.~70, pp. 258--265, 2016.

\bibitem{Cannon09m}
M.~Cannon, B.~Kouvaritakis, and X.~Wu, ``Model predictive control for systems
  with stochastic multiplicative uncertainty and probabilistic constraints,''
  \emph{Automatica}, vol.~45, no.~1, pp. 167--172, 2009.

\bibitem{Lazar2022}
M.~Lazar and P.~C.~N. Verheijen, ``Offset–free data–driven predictive
  control,'' in \emph{2022 IEEE 61st Conference on Decision and Control (CDC)},
  2022, pp. 1099--1104.

\bibitem{Doerfler2021}
F.~Dörfler, P.~Tesi, and C.~De~Persis, ``On the certainty-equivalence approach
  to direct data-driven {LQR} design,'' \emph{IEEE Transactions on Automatic
  Control}, vol.~68, no.~12, pp. 7989--7996, 2023.

\bibitem{maciejowski02predictive}
J.~M. Maciejowski, \emph{Predictive Control with Constraints}.\hskip 1em plus
  0.5em minus 0.4em\relax Pearson Education, 2002.

\bibitem{mosek}
\BIBentryALTinterwordspacing
M.~ApS, \emph{The {M}osek optimization toolbox for {J}ulia manual Version
  10.1.17.}, 2023. [Online]. Available:
  \url{https://docs.mosek.com/latest/juliaapi/index.html}
\BIBentrySTDinterwordspacing

\bibitem{waechter06}
A.~W{\"a}chter and L.~T. Biegler, ``On the implementation of an interior-point
  filter line-search algorithm for large-scale nonlinear programming,''
  \emph{Mathematical Programming}, vol. 106, no.~1, pp. 25--57, 2006.

\bibitem{Penrose1955}
R.~Penrose, ``A generalized inverse for matrices,'' \emph{Mathematical
  Proceedings of the Cambridge Philosophical Society}, vol.~51, no.~3, pp.
  406--413, 1955.

\end{thebibliography}

\end{document}